\documentclass[fleqn,usenatbib]{mnras}

\usepackage{newtxtext,newtxmath}

\usepackage[T1]{fontenc}
\usepackage{ae,aecompl}
\usepackage{afterpage}
\usepackage{adjustbox}
\usepackage{media9}
\usepackage[utf8]{inputenc}
\usepackage[T1]{fontenc}
\usepackage{parskip}

\usepackage{graphicx}	
\usepackage{amsmath}	
\usepackage{subfig}
\usepackage{comment}
\usepackage{enumerate}

\newcommand{\hMpc}{{\ifmmode{~h^{-1}{\rm Mpc}}\else{$h^{-1}$Mpc}\fi}}
\newcommand{\hGpc}{{\ifmmode{~h^{-1}{\rm Gpc}}\else{$h^{-1}$Gpc}\fi}}
\newcommand{\R}{{\ifmmode{~$R_{200}$}\else{$R_{200}$}\fi}}
\newcommand{\Msun}{{\ifmmode{{\rm {M_{\odot}}}}\else{${\rm{M_{\odot}}}$}\fi}}
\newcommand{\hMsun}{{\ifmmode{~h^{-1}{\rm M_{\odot}}}\else{$h^{-1}{\rm{M_{\odot}}}$}\fi}}

\newcommand{\threehundred}{\textsc{The ThreeHundred }}
\newcommand{\disperse}{\textsc{DisPerSE}}

\title[An inventory of filament galaxies]{An inventory of galaxies in cosmic filaments feeding galaxy clusters: galaxy groups, backsplash galaxies, and pristine galaxies}

\author[U. Kuchner et al.]{\parbox{\textwidth}{
Ulrike Kuchner,$^{1}$\thanks{E-mail: ulrike.kuchner@nottingham.ac.uk (UK)}
Roan Haggar,$^{1}$
Alfonso Arag\'{o}n-Salamanca,$^{1}$
Frazer R. Pearce,$^{1}$
Meghan E. Gray,$^{1}$
Agust\'{i}n Rost,$^{2}$
Weiguang Cui,$^{3}$
Alexander Knebe, $^{4,5,6}$
Gustavo Yepes$^{4,5}$
}
\vspace{0.4cm}
\\
\parbox{\textwidth}{
$^1$School of Physics \& Astronomy, University of Nottingham, Nottingham NG7 2RD, UK\\
$^2$Instituto de Astronomía Te\'orica y Experimental (IATE), Laprida 854, C\'ordoba, Argentina\\
$^{3}$Institute for Astronomy, University of Edinburgh, Royal Observatory, Edinburgh EH9 3HJ, United Kingdom\\
$^{4}$Departamento de F\'isica Te\'{o}rica, M\'{o}dulo 15, Facultad de Ciencias, Universidad Aut\'{o}noma de Madrid, 28049 Madrid, Spain
$^{5}$Centro de Investigaci\'{o}n Avanzada en F\'isica Fundamental (CIAFF), Facultad de Ciencias, Universidad Aut\'{o}noma de Madrid, 28049 Madrid, Spain
$^{6}$International Centre for Radio Astronomy Research, University of Western Australia, 35 Stirling Highway, Crawley, Western Australia 6009, Australia
}}

\date{Accepted 2021 November 21. Received 2021 November 19; in original form 2021 September 09}

\pubyear{2021}

\begin{document}
\label{firstpage}
\pagerange{\pageref{firstpage}--\pageref{lastpage}}
\maketitle

\begin{abstract}
Galaxy clusters grow by accreting galaxies from the field and along filaments of the cosmic web. As galaxies are accreted they are affected by their local environment before they enter (pre-processing), and traverse the cluster potential. 
Observations that aim to constrain pre-processing are challenging to interpret because filaments comprise a heterogeneous range of environments including groups of galaxies embedded within them and backsplash galaxies that contain a record of their previous passage through the cluster.
This motivates using modern cosmological 
simulations to dissect the population of galaxies found in filaments that are feeding clusters, to better understand their history, and aid the interpretation of observations. 
We use zoom-in simulations from \threehundred project to track halos through time and identify their environment. We establish a benchmark for galaxies in cluster infall regions that supports the reconstruction of the different modes of pre-processing. 
We find that up to 45\% of all galaxies fall into clusters via filaments (closer than 1$ \hMpc$ from the filament spine). 12\% of these filament galaxies are long-established members of groups and between 30 and 60\% of filament galaxies at \R\ are backsplash galaxies. This number depends on the cluster's dynamical state and sharply drops with distance. Backsplash galaxies return to clusters after deflecting widely from their entry trajectory, especially in relaxed clusters. They do not have a preferential location with respect to filaments and cannot collapse to form filaments. The remaining pristine galaxies ($\sim$30 -- 60\%) are environmentally effected by cosmic filaments alone.
\medskip


\end{abstract}



\section{Introduction}
\label{sec:introduction}

In a cold dark matter Universe model, low-mass halos form first. Larger halos become more common over time, successively building up their mass through merging and accretion of smaller halos. Galaxy clusters mark the culmination of mass assembly and the peaks of dynamical gravitational structure formation. They are the highest density environments in the large-scale Universe, packed with thousands of galaxies, both in the virialized cluster core and infalling from the highly anisotropic matter distribution surrounding the clusters. Galaxies fall into clusters through a variety of environments: as part of groups, on their own from the general field, or in streams via filaments of the cosmic web \citep{Zeldovich1970, Haarlem1993}. 

Clusters assemble their \textit{mass} predominantly by massive accretion events, like infalling groups of galaxies \citep{McGee2009}, but build up their \textit{galaxy} population predominantly through the accretion of lower mass halos, i.e., isolated galaxies falling in on their own \citep{Berrier2008}. 
This differentiation could be important for the evolution of galaxies and clusters of galaxies, because different environments evoke different physical processes that depend on the mass of the host (dark matter) halo. 
Satellite galaxies in high density environments such as clusters and groups differ from isolated galaxies of the same stellar mass in key aspects, such as their colour \citep[e.g.,][]{Peng2010}, star formation rate \citep[e.g.,][]{kauffmann04, wetzel13}, and morphology \citep[e.g.,][]{Dressler1980}. Galaxies in denser environments tend to be redder, more elliptical/spheroidal with less gas and ongoing star formation. 
This well-known finding is grounded on a wealth of observations from galaxies in clusters opposed to galaxies in the general field and tested against a variety of physical processes acting in clusters \citep{Oemler1974, Dressler1980, postman84, Balogh1997, Poggianti1999}. Galaxies are commonly thought to transform both in terms of star forming activity and morphology as they experience dense environments. Therefore, the environment of galaxies plays a key role in the formation and evolution of galaxies \citep{Blanton2009}. Still, a full description of the relationship between galaxies and their environments, including their specific processes (i.e., environmentally driven tidal or hydronamical mechanisms versus internal mass-dependent mechanisms) is still outstanding. A primary complication for an understanding is that we do not know how much of the correlation between galaxy properties and cluster membership is due to a transformation inside the cluster as opposed to in environments prior to entering the cluster, a phenomenon called "pre-processing" \citep{Zabludoff1998, balogh00, wetzel13}. While this term is not absolute, it generally refers to any process operating in high density environments that leads to the transformation of galaxies and experienced before the cluster infall. In this paper, "pre-processing" therefore summarises \textit{all} environmental effects, including hydrodynamical and gravitaional effects, acting in groups and in large-scale filaments, that affect cluster galaxies before they enter the virialized regions of a cluster. A common indicator for pre-processing is galaxy quenching, as this presents relatively accessible observational evidence of pre-processing, however changes in other galaxy properties like galaxy structures can equally help to constrain pre-processing.

The increasing awareness and current discussion of pre-processing as an important ingredient to galaxy evolution has prompted surveys to focus on cluster outskirts, i.e., observations that go beyond \R\footnote{The radius within which the mean density of a cluster is equal to 200 times the critical density of the Universe and used by us as defining the extend of the cluster.}, in order to identify the sites where galaxies are first affected by their environment before falling into clusters \citep{Fujita2004, Porter2008, Mahajan2012, Haines2018, Sarron2019, Malavasi2020}. One upcoming dedicated study of cluster infall regions is the WEAVE Wide-Field Cluster Survey (WWFCS) with the multi-object spectrograph WEAVE (WHT Enhanced Area Velocity Explorer) on the William Herschel telescope \citep{Balcells2010, Dalton2012}. It will systematically observe 20 nearby clusters out to 5\R\ with the goal to determine whether significant pre-processing accelerates the quenching of star formation and/or morphological transformation. Our investigation presented in this and previous papers is motivated by the WWFCS, but the results are universal and equally applicable to a wide range of experiments.

In this paper, we focus on pre-processing in large-scale filaments, which themselves are heterogeneous environments, including galaxy groups embedded within them. Around half of the mass of the Universe is found in cosmic filaments \citep{Cautun2014, Cui2018}, which, in turn, fundamentally define the spatial organisation of galaxies over a vast range of scales from less than one to tens and even hundreds of Megaparsecs \citep{Libeskind2017, vandeWeygeartProceedings2014}.
A growing body of evidence shows that large-scale filaments play a similar role in shaping the properties of galaxies as clusters do, albeit to a lesser degree. Galaxies close to cosmic web filaments are redder \citep{Kraljic2018, Laigle2017}, elliptical \citep{Kuutma2017}, with higher metalliciy \citep{Darvish2015, Gray2009}, more massive \citep{Malavasi2016} and more likely to have been quenched \citep{Alpaslan2016, Winkel2021} than their counterparts at fixed M* at increased distances away from filaments. This can be due to ram pressure that removes the hot halos especially of lower mass galaxies \citep{Bahe2013,Benitez-Llambay2013}. 
While simulations suggest that halos at the same mass in denser environments form earlier than in less dense environments, owing to the dependence of halo clustering not only on mass but also on the formation redshift and assembly history \citep[a term coined "assembly bias",][]{Gao2005, Jung2014}, this may be a simplified view of the problem since mass assembly is driven by  different physical processes inside and outside of filaments \citep{Poudel2017}. Mergers, tidal effects and smooth accretion are attributed to different densities and strongly influence the current property of a galaxy beyond its formation time. Differences can also be explained by accretion of pre-enriched filamentary gas \citep{Darvish2015}, which may lead to a star-formation enhancement in filaments \citep{Vulcani2019} when galaxies are fuelled with gas \citep{Kleiner2016}. 

Embedded within the large-scale cosmic web, galaxy groups continue to accrete galaxies and gas \citep{Kauffmann2010}. This is especially relevant close to clusters, where infalling groups can easily sweep up field galaxies and grow quickly \citep{Vijayaraghavan2013}. Members are likely processed by ram pressure enhanced by feedback within groups prior to their accretion into the clusters themselves \citep{Bahe2014, Jung2018}, as is evident in observations of galaxy mergers and ram pressure stripping signatures \citep{Jaffe2016, Bianconi2017, Haines2018a, Benavides2020}. Earlier simulations suggest that a significant fraction of all cluster galaxies -- some report between a third and half of cluster galaxies at $z=0$ --  could enter clusters as part of groups \citep{McGee2009,White2010,DeLucia2012}. However, most galaxies spend relatively little time in groups before falling into the cluster \citep[less than 2.5 Gyrs,][]{ Vijayaraghavan2013, Han2018}, so either group environmental mechanisms must act fast to be significant for the cluster population, or only group members that have spent extended periods of time in their host halo are measurably affected and indeed pre-processed. Either way, most groups are part of filaments \citep[e.g.,][]{Tempel2014}, and therefore a number of filament galaxies are actually processed by their group environment. To unambiguously identify the effect of filaments on galaxy evolution, it may be necessary to remove the contribution of groups. 

After galaxies are accreted by the cluster, they either remain bound to the gravitational potential well of the cluster, or their trajectories carry them out of the cluster, up to several \R, where they will turn around to fall back in on a subsequent infall. This population of "backsplash galaxies" is no small fraction: immediately outside of clusters, up to 70\% of all galaxies can be backsplash galaxies \citep{Gill2005,Haines2015, Haggar2020} and have therefore been processed by the cluster itself. By the time they are observed as backsplash galaxies, they may reveal their past environmental history through "post-processing" signatures that are all but indistinguishable to pre-processing signatures. Beyond this complication, other possible processing mechanisms induced in accretion shocks or when crossing cosmic web walls  
\citep["wall stripping", ][]{Winkel2021} can strip halo gas which leads to star formation consumption and quenching, especially in low-mass galaxies. 

As a direct consequence of structure formation, galaxies falling into clusters are therefore a combination of "field galaxies" -- both isolated and as pairs and small groups -- and galaxies in filaments -- again, isolated and as part of groups -- as well as backsplash galaxies.
Given this diversity, pre-processing studies need to take the entire environmental history of galaxies over a lifetime spent in a hierarchically assembling global environment into account.
This paper sets out to provide a census of the fractions of galaxies that feed clusters from a variety of evolving environments and investigates whether this varying composition depends on the dynamical state or mass of the cluster. This can inform analysis of observational signatures of star formation histories against measured environments which investigate galaxy transformation.
Our study involves tracing the filamentary structure beyond the virial radius in large hydrodynamical simulations while also considering the orbital trajectories of infalling galaxies. After detailing the simulations (Sec. \ref{sec:300}), we discuss the identification of the main components, filaments(Sec. \ref{subsec:filament_identification}), groups (Sec. \ref{subsec:group_identification}) and backsplash galaxies (Sec. \ref{subsec:backsplash_identification}). We then discuss the importance of group galaxies and filaments (Sec. \ref{sec:groups_importance}) and the contamination of backsplash galaxies in filaments, where we separate galaxies that are leaving the cluster from returning galaxies (Sec. \ref{subsec:backsplash_in_filaments}). Our final section summarizes the heterogeneous composition of filament galaxies.  

\section{Simulations and Methods}
\label{sec:simulations_methods}

\subsection{\threehundred clusters}
\label{sec:300}
To help interpret observations of galaxy properties feeding clusters, we need to know their environmental history during accretion. To know this means to turn to simulations. 
This paper makes use of hydrodynamical simulations of \threehundred project\footnote{https://the300-project.org} \citep{Cui2018}. This project selected 324 spherical regions with radius $15\hMpc$ centered on the most massive clusters ($M_{200} \gtrsim \times 10^{14} \hMsun$) in the $1 \hGpc$ volume of the dark-matter-only MultiDark simulation \citep{Klypin2016} with Planck cosmology \citep{Ade2016}. \threehundred were simulated using a range of different physics models. The suite contains the same clusters simulated with Gadget-Music \citep[][]{Sembolini2012}, Gadget-X \citep[][]{Beck2015, Rasia2015} and GIZMO-Simba \citep[][]{Dave2019} amongst others, as well as several semi-analytic models, producing 129 snapshots from redshift $z \sim 17$ to $0$. For a comprehensive description and discussion of the full-physics treatment, comparison and limitations of codes and the \textsc{AHF}-halo finding of \textsc{The ThreeHundred}, we refer to the survey description paper by \citet{Cui2018} and references therein. For the work presented in this paper, we only use the mass distribution of the full physics simulations performed with Gadget-X to generate our filamentary network, because the goal of this investigation does not require further information. With the exception of tracing the infall of galaxies to identify backsplash galaxies, we restrict our current investigation to redshift z=0, both motivated by the wish to minimize evolutionary effects, and preparing for upcoming observations with WEAVE. We will expand on this in a future publication (Cornwell et al. in prep).

In summary, these hydrodynamic simulations of galaxy clusters return information in 6 dimensional phase space over numerous time steps in a volume of several virial radii of the clusters, i.e., large enough to include many additional groups and filaments, which may or may not be physically associated with the central cluster and useful to track infall. The sample also includes volumes that host pairs of clusters.
We assess the dynamical state of the cluster, "relaxedness", based on a combination of three characteristic parameters that capture signatures of activity. These are 1) the virial ratio (a measure of virialization of the cluster), 2) the centre-of-mass offset from the maximum density point, and 3) the fraction of mass in subhalos \citep[see][where this has been discussed in detail]{Cui2017, Cui2018, Haggar2020}. A cluster is considered relaxed if it has a low fraction of mass in subhalos, low centre-of-mass offset and virial ratio approaching 1. Clusters with a higher "relaxedness" parameter $R$ and specifically with $R>1$ are considered more relaxed, clusters with $R<1$ as unrelaxed or dynamically active.

How reliably we can separate galaxies in groups, filaments or cluster outskirts is fundamental for studying the effects of galaxy evolution and pre-processing. Systematics in classifications can bias our view of pre-processing and hamper the compatibility of simulations and observations. Simulations can help to quantify the effect of every environment a galaxy experiences during its lifetime but some care needs to be taken to bridge simulations to observations. In previous publications, we have detailed how transforming the simulations into realistic mock observations allow to forecast the impact of projection effects and the reliability of filament finding for upcoming wide-field spectroscopic surveys \citep{Kuchner2020, Kuchner2021}. In a next step, we will investigate the effects of further observational constraints such as fibre collisions during the production of observing blocks on finding filaments in the crowded regions of galaxy clusters (Cornwell et al in prep.).
While we especially focus on mimicking observations that will be obtained with the WEAVE Wide-Field Cluster survey as part of the community-led surveys with the new spectroscopic facility WEAVE at the WHT (see Introduction Sec. \ref{sec:introduction}, as well as \citet{Kuchner2020} and Jin et al in prep.), we emphasise that results are more general, and valid for a number of observational applications.  
\begin{figure}
	\includegraphics[width=\columnwidth]{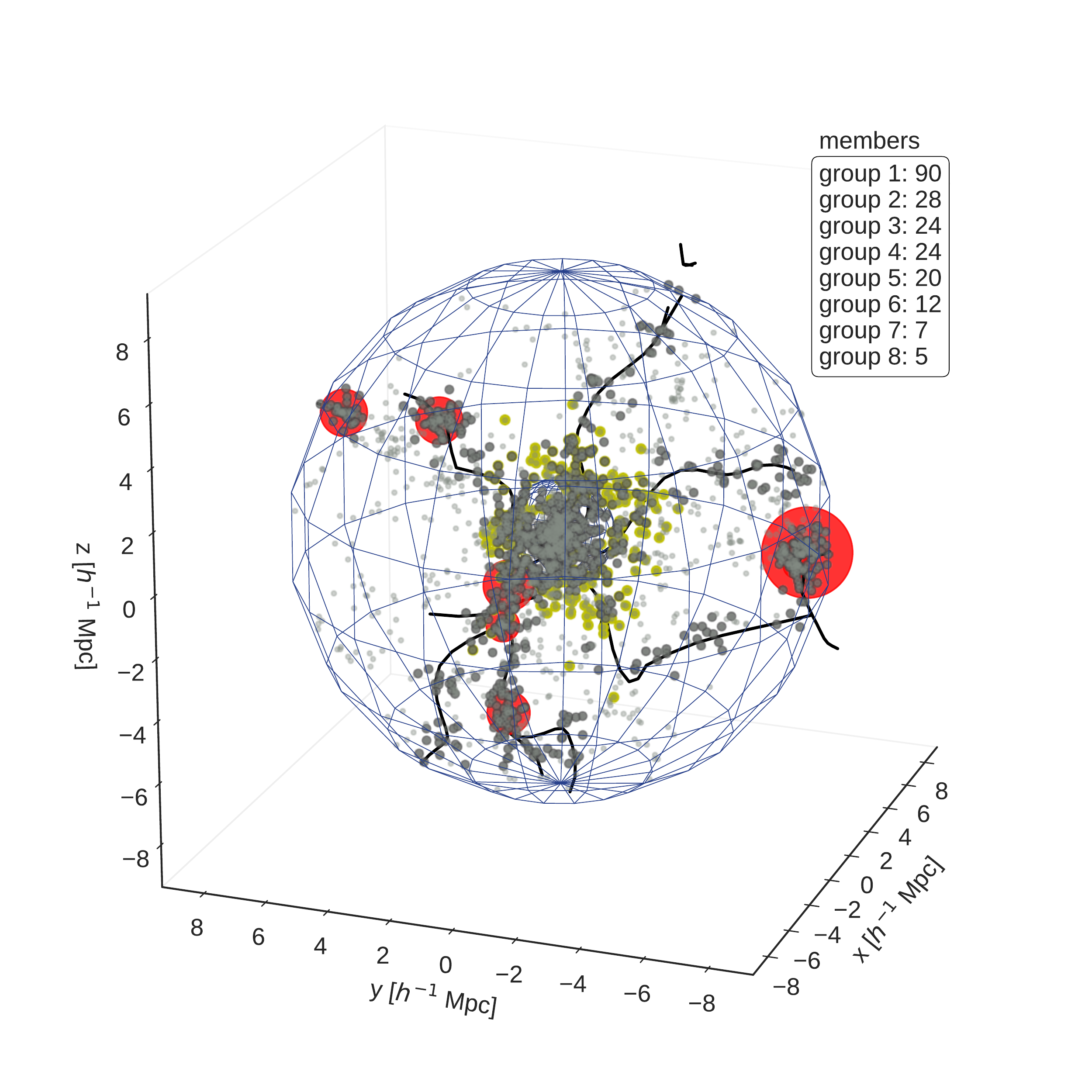}
    \caption{
    The example cluster (cluster 0066) of \threehundred\ project illustrates the variety of environments and processing histories galaxies around clusters can have. Shown are galaxies within $1\hMpc$ of galaxy-detected filaments (dark gray), groups (highlighted by red disks), the general "field" (light gray) and backsplash galaxies (yellow).  Each environment is related to mechanisms that may pre-process the galaxies. Backsplash galaxies have been environmentally affected by the cluster itself on their previous pass through the cluster. The large mesh sphere indicates 5\R, the small sphere 1\R. The insert lists the number of group members for this example.}
    \label{fig:groups}
\end{figure}

\subsection{Filament identification}
\label{subsec:filament_identification}

The paper considers major filaments around clusters that can be thought of as \textit{highways} or \textit{transport channels} of the Universe (Fig. \ref{fig:groups}), along which mass and galaxies get funneled into clusters \citep[e.g.,][]{Haarlem1993,Knebe2004}. To extract cosmic web filaments in each volume of our sample at $z = 0$, we used the robust filament finding algorithm \disperse\ \citep{Sousbie2011}. We have applied the software on a discrete point distribution of mock galaxies in 3D and 2D -- a useful and well established approach in both simulated and observed datasets across scales from sub-galactic to cosmological interests \citep[e.g.,][]{Malavasi2016, Malavasi2020, Kraljic2018, Hess2018, Arzoumanian2019, Winkel2021}. For our purpose, we define mock galaxies as all halos with masses $M_{\rm{halo}} > 3\times 10^{10} \hMsun$ (comparable to $M_{\rm{*}} > 3\times 10^{9} \hMsun$ \footnote{In \citet{Kuchner2020}, we discussed how halo mass limits compare to stellar masses expected for upcoming WWFCS observations that motivate this choice. In the present paper, we continue to use halo masses.}) and use them as input to \disperse. 
The software processes the data in two steps. For filaments used in this paper, the software first computes the density from the Delaunay Tesselation on the 3D halo distribution, which we post-process by weighting by halo mass of each mock galaxy. Then, \disperse\ identifies the critical points in the density field; in 3D, these are minima, two kind of saddle points, and maxima. The filament extraction is theoretically motivated: filaments are defined as the spatial lines following the gradients of the density field that connect saddle points to peaks. Not all critical points that \disperse\ extracts have the same significance with respect to noise. The significance of a pair of critical points (e.g., a saddle-to-peak pair) with respect to the noise is quantified by the persistence parameter $\sigma$, which is a user-controlled input parameter. 
This way, low persistence features can be filtered out, which in turn allows to work with noisy data sets and to remove features that are not physically meaningful. In \citet{Kuchner2020} we compared filament networks based on mock galaxies in true 3D coordinates to the networks based on the underlying gas distribution, which we considered as our reference network. The result of this assessment was a persistence threshold of $\sigma = 6.5$ appropriate for finding filaments based on mock galaxies around massive clusters. 

The output of the algorithm is a set of critical points and spatial lines presented as small segments of the filament axes (or skeleton). We can therefore compute the distance of each mock galaxy to the filament axes, a useful parameter to investigate gradients of galaxy properties \citep[e.g.,][]{Laigle2017, Kraljic2018}. \disperse\ does not give information whether a galaxy is "inside" or "outside" a filament. In order to compute the filament diameter or width, an additional parametrisation is required. In \citet{Kuchner2020}, we have defined the filament width based on density profiles of gas particles as a way to provide a convenient "inside/outside" definition for observational applications. We have taken care to choose an appropriate fixed width, trying to optimise completeness without increasing the contamination. Depending on the science goal, we defined mock galaxies with distances to filament axes (skeleton) of $D_{\rm{skel}}<0.7 \hMpc$ (for maximum purity) or $D_{\rm{skel}}<1 \hMpc$ (for maximum completeness) to be "inside" filaments. Note that a constant thickness and basic segregation is a simplification that does not properly reflect the diffuse characteristic of filament gas and galaxies collapsing towards filament spines, nor does it properly capture the variation of filament thickness closer to halos including at locations of massive groups \citep{Dolag2006, Rost2020a}. 

In this paper, we define filaments with a constant thickness of $1 \hMpc$, i.e., all mock galaxies with a distance of less than $1 \hMpc$ to the skeleton ($D_{\rm{skel}} < 1 \hMpc$) are considered filament galaxies (Fig. \ref{fig:groups}). This is similar to choices made in previous publications \citep[e.g.,][]{Colberg2005, Tempel2014, Kooistra2019}. We also note values for a more restricted filament thickness of $0.7 \hMpc$ in the text. All numbers thus depend on the choice of filament thickness, which in turn depends on the science case and emphasis on e.g., purity vs. completeness.
The density profile discussed in Fig. 6 in \citet{Kuchner2020} shows that the profile drops steeply beyond 
$1 \hMpc$. Increasing the filament thickness by a factor of two ($D_{\rm{skel}}<2 \hMpc$) therefore leads to a large increase of contamination 
while overall only adding $\sim4$ per cent of galaxies that are located in the true periphery of filaments.
Importantly, it is not clear whether these galaxies will experience any environmental effect in filament peripheries, since at $2 \hMpc$ from the filament spine, the density has dropped by a factor of $\sim12$ (depending on proximity to the node), which will be difficult to verify observationally. In summary, the choice of a constant and unique thickness for all filaments remains a simplification and does not fully capture the variation in filaments, but it considers the majority of true filament galaxies that experience a significant increase of gas density while keeping the contamination at bay.

 

\subsection{Group identification}
\label{subsec:group_identification}
Our group identification is motivated by observations and the overall objective to identify group members that can experience pre-processing. Finding groups in observations is a challenging problem, because, while groups comprise all gravitationally bound galaxies residing in a dark matter halo, often only the brightest (usually central) galaxy or galaxies can be detected due to the survey's magnitude limit. Background and foreground objects and redshift space distortions lead to high false positive rates.
In that case, one might choose to first identify bright group galaxies based on their spectroscopic or line-of-sight velocity data. Then, an excess of fainter galaxies in comparison to a field sample can be assigned to the group.
Alternatively, a number of automated ways (geometrical, colour and model-based methods as well as probabilistic techniques) to identify galaxy agglomerations in large-scale survey observations exist, including methods like the Dressler–Shectman tests \citep[DS;][]{Dressler1988}, halo-based group finders \citep[e.g.,][using halo occupation statistics]{Yang2005}, Voronoi-Delaunay Method \citep{Marinoni2002} and Friends-of-Friends algorithm \citep{Geller1983} or through X-ray observations that bypass the uncertainty from small numbers of luminous galaxies in groups. Each recipe to find group members comes with benefits and drawbacks, and fair comparisons are understandably challenging.
If spectroscopic data is available, groups in and around clusters specifically have often been identified using positions and velocities \citep[e.g.,][]{Eke2004, Lisker2018, Iodice2019}. The aim is to select galaxies that most likely represent the true bound structures, however, science-specific considerations (e.g., completeness versus purity) will control choices.

Similar to this idea, we define group galaxies in \threehundred simulations by first locating group centre halos outside of 1$R_{200}$ and within 5$R_{200}$ of the central cluster. These are halos with velocity dispersion $\sigma_v > 300\ h^{-1}\rm{km/s}$ 
and mimic the most luminous central galaxy of the group. For reference, this is slightly higher than the median velocity dispersion of groups in the Two-degree Field Galaxy Redshift Survey  \citep[2dFGRS, ][]{Eke2004}. Then, we identify group members as all mock galaxies (i.e., halos above $\sim 3\times 10^{9} \hMsun$) within 1$R_{200}$ of this central halo. An additional criterion based on the distance to the group centre assures that each halo can only be a member of one group. Note that by lowering the velocity dispersion threshold to $150 \ h^{-1}\rm{km/s}$, many more galaxies in groups can be identified, and the fraction of galaxies that are members of groups almost doubles. However, we prefer to select group members with a higher probability to have been affected by the high density environment, i.e., galaxies that have experienced pre-processing.
This is because close to clusters, infalling groups can easily sweep up field galaxies and grow quickly \citep{Vijayaraghavan2013}. Therefore, while larger fractions of galaxies may enter the cluster through lose groups or pairs, most have only had a brief pre-processing period \citep{Han2018}.
Our approach does not exclude very rich sub-structures that could be considered as discrete clusters (see Sec. \ref{subsec:scatter}). We do not impose a sharp divide between a group and a cluster since observationally, numbers of members depend on further quantities like magnitude- or volume limits.  Furthermore, group definitions span a variety of properties like size and  richness, with a wide range of velocities and morphologies of its members.

Fig. \ref{fig:groups} shows an example of groups highlighted in red identified in one cluster volume. The insert prints the number of group members for each group: the largest group found in this volume has 90 galaxies, the smallest 5.
The figure also highlights filaments in black (Sec. \ref{subsec:filament_identification}) with associated filament galaxies 
in dark grey and backsplash galaxies (see next Sec. \ref{subsec:backsplash_identification}) in yellow. Because filaments can be understood as ridges that connect maxima (nodes) in the density field of the galaxy distribution, we see filaments linking groups and clusters. Therefore, most group members will be part of the filament network, located in the cores of filaments. Likewise, filament galaxies, as defined by their distance to the skeleton ($D_{\rm{skel}} < 1 \hMpc$), can be group members. The simplification of a fixed filament width also means that some group members of massive groups will be located further than $1 \hMpc$ from the filament spine.
\begin{figure}
  \centering
  \subfloat[dependence on cluster mass]{\includegraphics[width=0.47\textwidth]{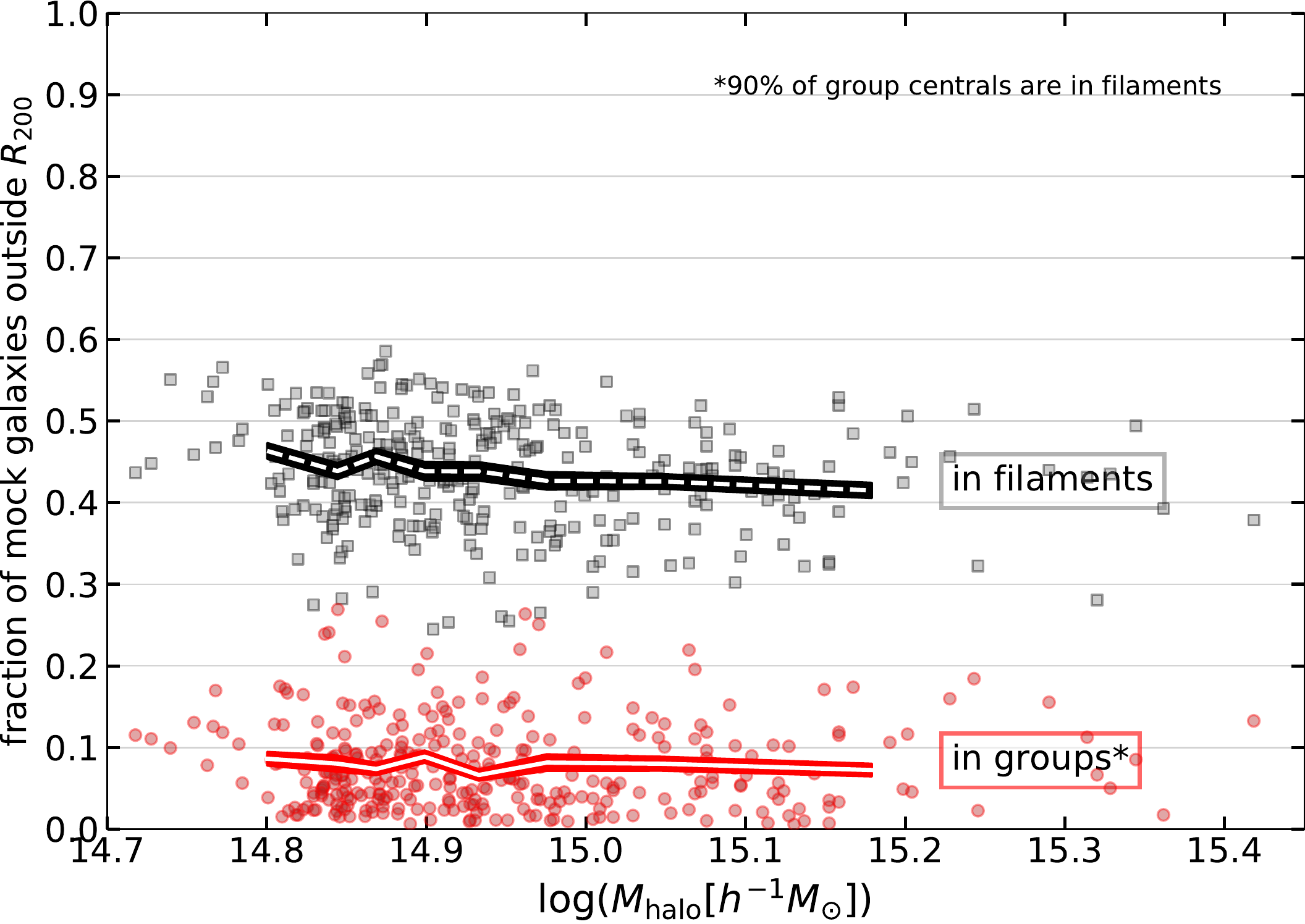}\label{fig:groups_mass}}
  \hfill
  \subfloat[dependence on dynamical state of the cluster]{\includegraphics[width=0.47\textwidth]{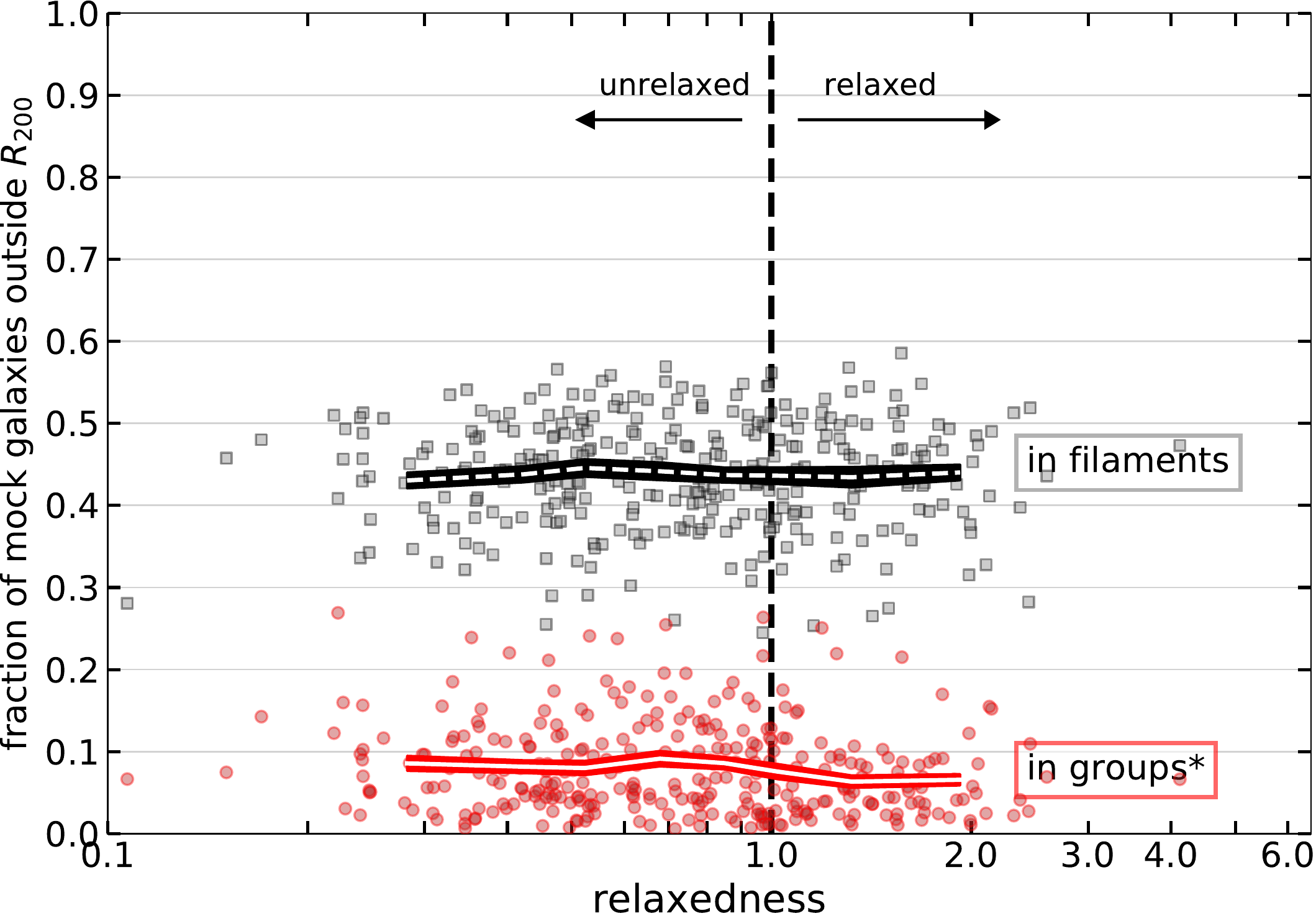}
  \label{fig:groups_relaxedness}}
  \caption{The fraction of all galaxies in filaments (black band, dashed line, defined as halos with distance to filament axes $D_{\rm{skel}}<1 \hMpc$) and in groups (red band, solid line) outside \R\ does not depend on mass (top) or dynamical state (bottom) of the cluster. About 45 per cent all galaxies down to $M_* >10^9 M_{\odot}$ are in filaments, and $\sim$10 per cent of all galaxies are in groups. The number of filament galaxies depends on the choice of filament thickness. Here we consider filament cores with a constant radius of $1 \hMpc$. The fractions are not exclusive: 90 per cent of central group halos are part of filaments and therefore a large fraction of group galaxies are also in filaments. Unrelaxed clusters are roughly defined as cluster with relaxedness $R<1$. Coloured bands are 1$\sigma$ errors on the mean. }
  \label{fig:fractions}
\end{figure}
\begin{figure*}
	\includegraphics[width=0.95\textwidth]{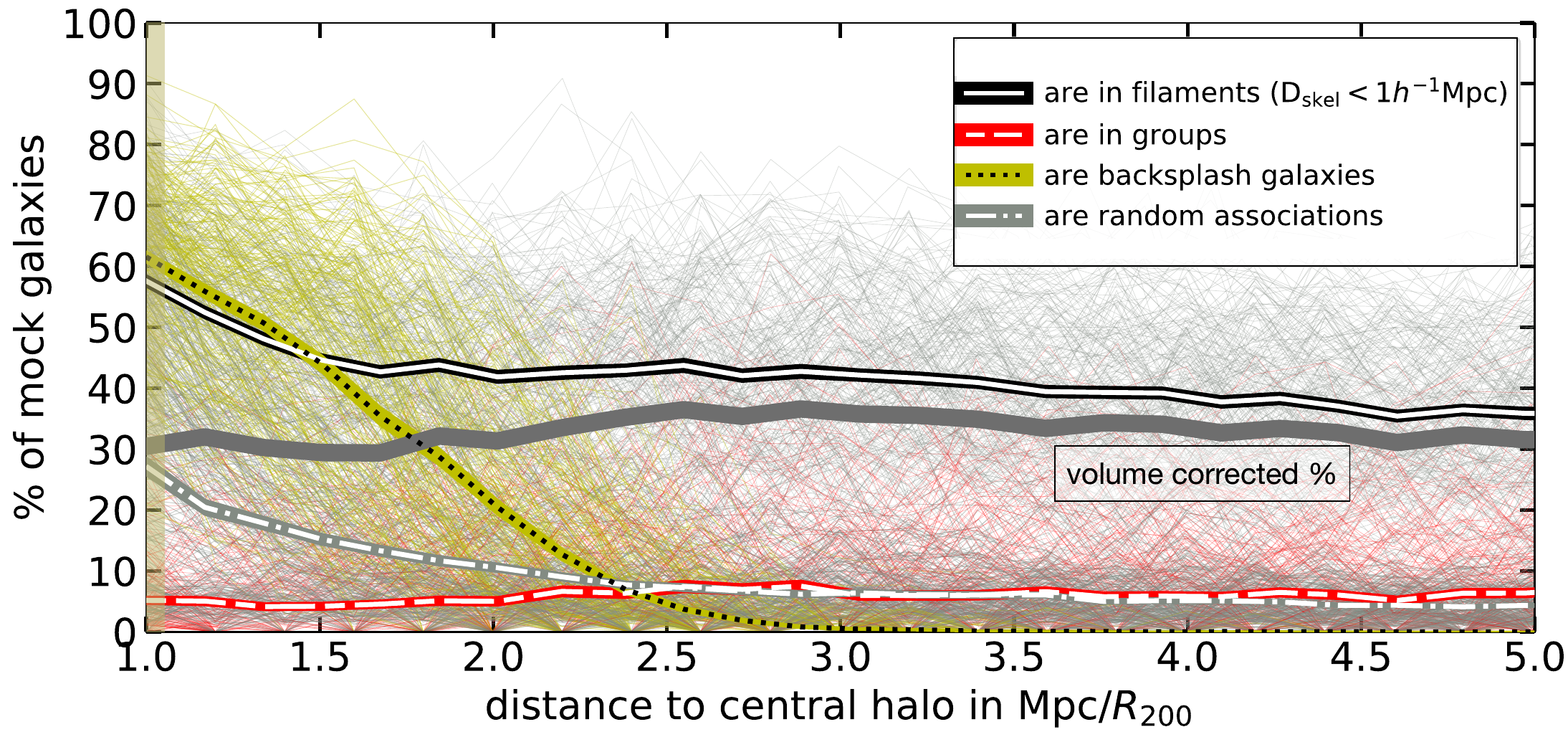}
    \caption{The fraction of galaxies in groups and filaments does not depend on the cluster distance, backsplash galaxies increase steeply closer to the cluster. Shown are percentages of mock galaxies in filaments (black  band, solid line), groups (red band, dashed line) and that are backsplash (yellow band, dotted line) as a function of distance to the cluster centre (normalized by \R). Note that the number of filament galaxies is for a characteristic filament thickness of 1Mpc ($D_{\rm{skel}}<1 \hMpc$) and bands are the 1$\sigma$ error on the mean. The fractions are not exclusive. The thick solid dark grey line is the volume-corrected number of galaxies in filaments based on a randomly rotated networks (light grey dot-dashed line).}
    \label{fig:frac_distance_all}
\end{figure*}

\subsection{Backsplash galaxy identification}
\label{subsec:backsplash_identification}

In the most general terms, backsplash galaxies are galaxies that are observed outside \R\ of the cluster, but have been inside of the cluster previously \citep{Gill2005, Bahe2013}. As a result, these galaxies have likely undergone significant disruption. They can be either departing (leaving) the cluster after its passage through, or they area on a subsequent infall (returning). This definition does not assume that the galaxy is bound to the cluster halo and does not include the location of the galaxy outside the cluster. Note that this definition is not unique\footnote{E.g., \citet{Haines2015} consider all galaxies on their outward radial velocity past pericenter as backsplash galaxies.}. While they are found in the same location as infalling galaxies, and are thus only distinguishable through kinematics for the observer \citep{Gill2005, Pimbblet2010}, backsplash galaxies have been affected by the cluster environment itself. 

In \threehundred simulations, we identify backsplash galaxies based on the orbital history of each galaxy relative to \R. The backsplash galaxy population consists of all galaxies with a distance to the cluster center at $z=0$ of $D_{z=0} > R_{200}$ and a minimum distance to the cluster centre at any time in their history $D_{min} < R_{200}$. 
For an analysis that includes backsplash galaxies -- which require knowledge of previous snapshots $z>z_0$ -- we use a subsample of 257 clusters. Briefly, clusters and their backsplash population are excluded from the sample in cases when the main branch cannot be tracked back to before $z = 0.5 $ and when large apparent jumps in the position of the cluster merit a judgement on $D_{z=0} > R_{200}$ unreliable \citep[see][for details]{Haggar2020}. Fig. \ref{fig:groups} shows backsplash galaxies in yellow: their distribution forms a cloud around the clusters' \R. Note that affiliation to the backsplash population and group membership are not exclusive. Backsplash galaxies can be part of groups, however with only 9\% of backsplash galaxies in groups, this is relatively rare.

\section{Results and Discussion}
\subsection{The importance of galaxies in groups and filaments surrounding clusters}
\label{sec:groups_importance}

Galaxies that are part of groups and filaments prior to the final cluster environment may have been environmentally affected, i.e., pre-processed. Because of their sufficiently high densities, but lower velocity dispersions (and therefore higher dynamical friction force), transformation and merging occur more frequently in groups than in clusters. Therefore, in order to understand the role of pre-processing, it is important to know how many infalling galaxies are part of groups. Furthermore, groups are usually part of the wider filamentary network, as they represent maxima in the density field. Thus, most group galaxies are also filament galaxies. 
\begin{figure*}

\centering

\subfloat[low contamination]{%
  \includegraphics[width=0.31\textwidth]{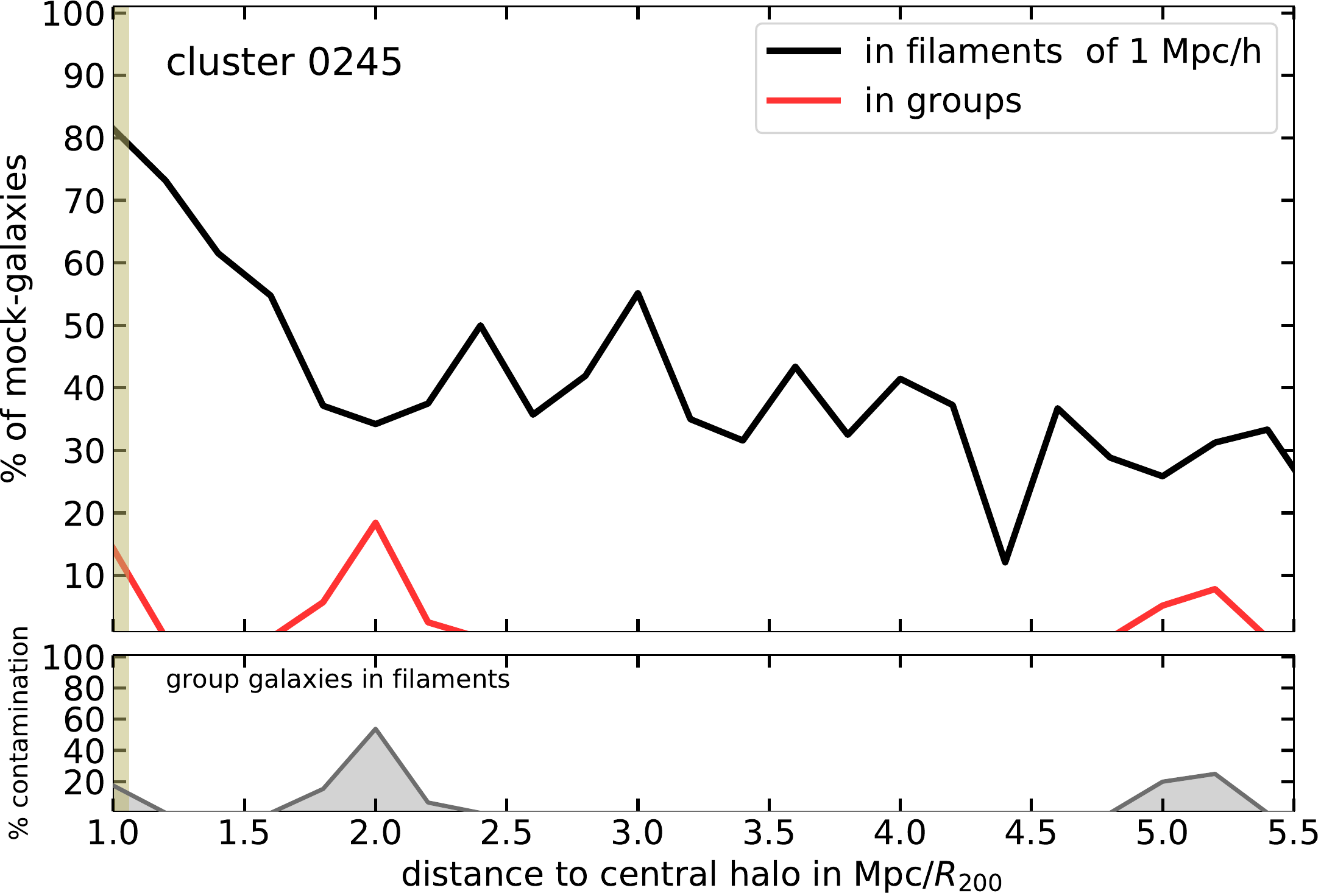}%
  \label{fig:scatter1}%
}\qquad
\subfloat[medium contamination]{%
  \includegraphics[width=0.31\textwidth]{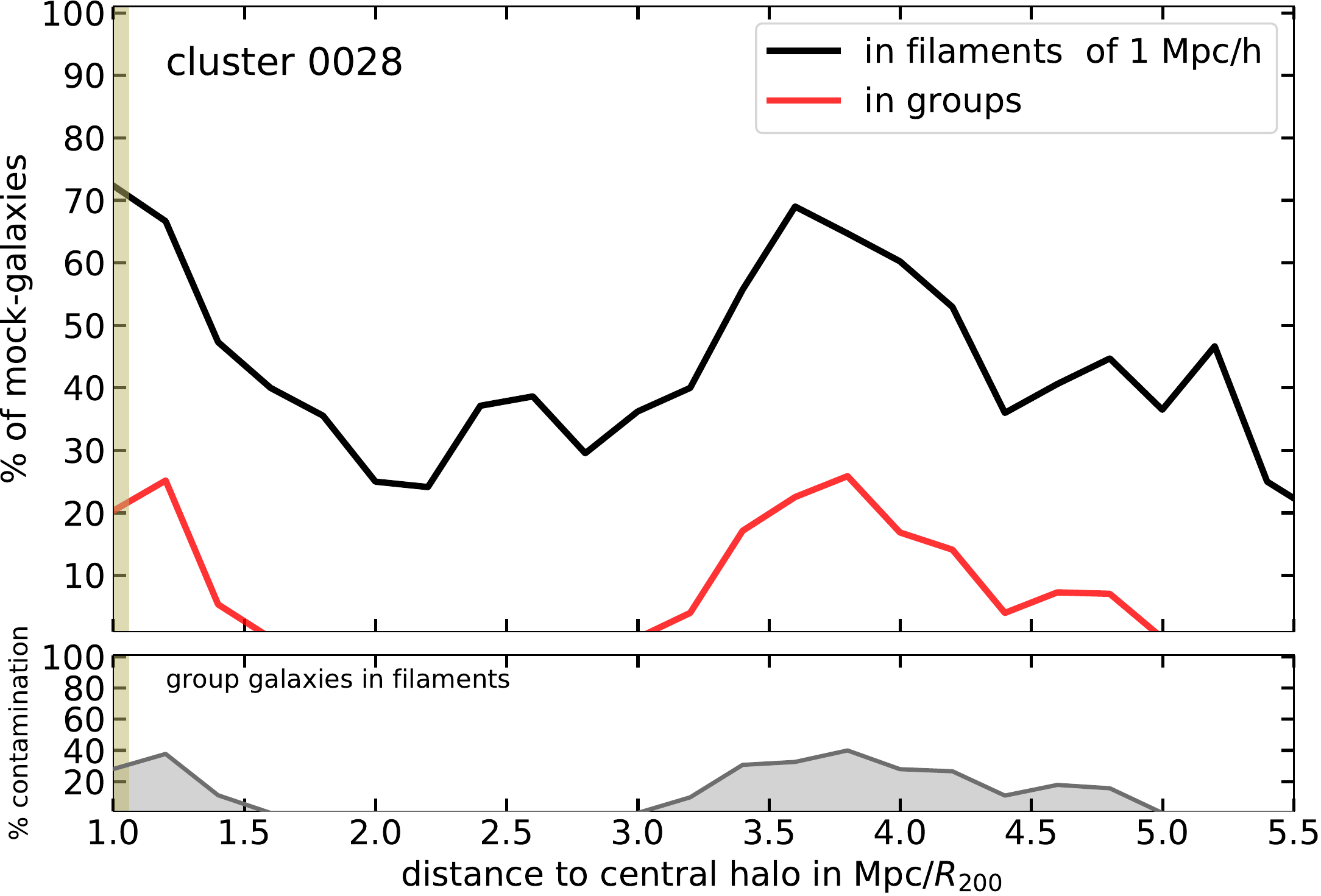}%
  \label{fig:scatter2}%
  }\qquad
\subfloat[high contamination]{%
  \includegraphics[width=0.31\textwidth]{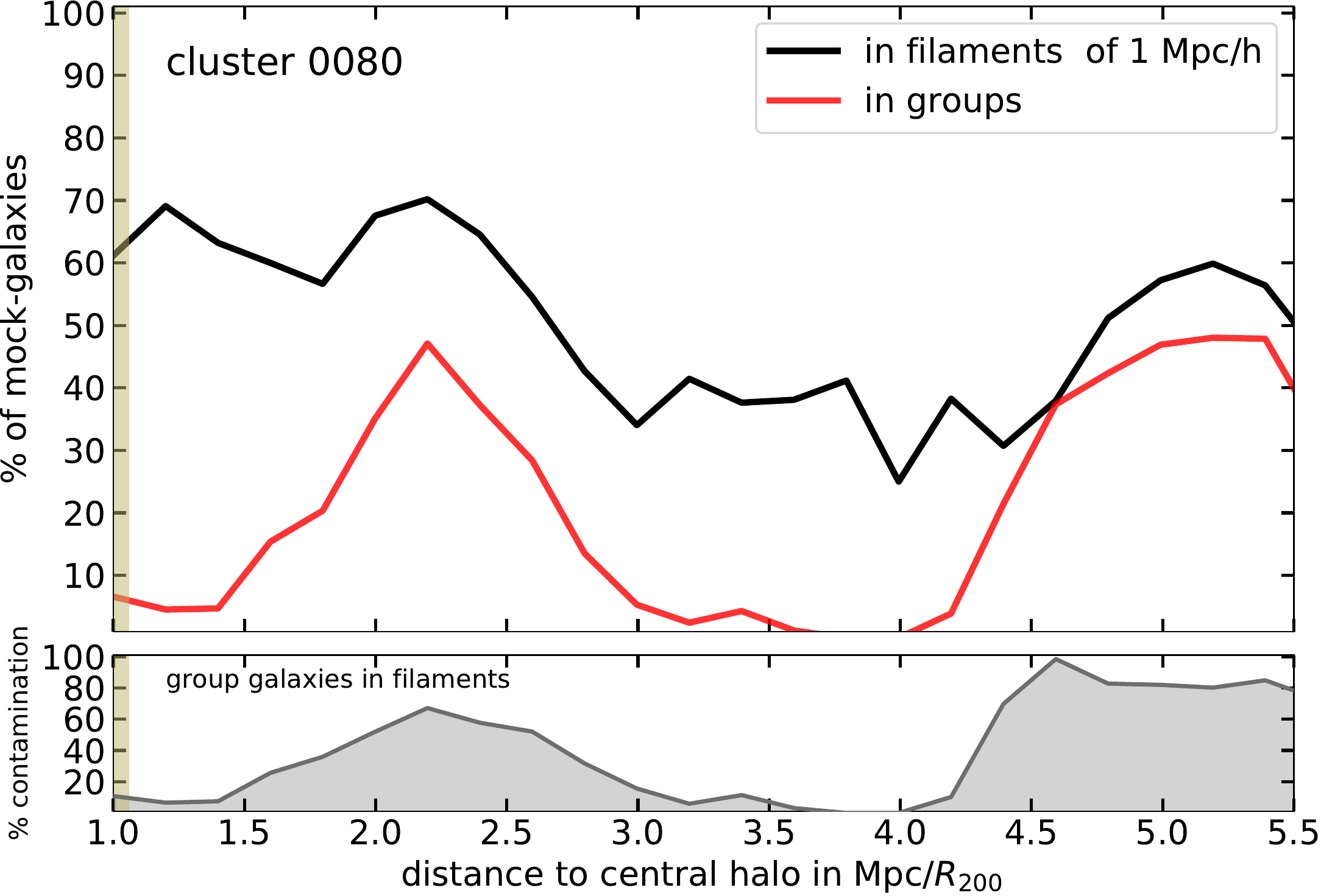}%
  \label{fig:scatter3}%
}
\caption{Three examples of varying contamination of group galaxies in filaments of $1 \hMpc$ thickness feeding galaxy clusters. It shows distances to the cluster centre versus the fraction of galaxies in filaments (in black) and galaxies in groups (in red). The peaks in the percentage of mock galaxies indicate the positions of the groups and also shows that the groups are part of filaments. The extreme example on the right is a complex system including a cluster-sized group of 180 members.}
\label{fig:scatter}
\end{figure*}

We therefore ask: are the filaments feeding galaxy clusters dominated by distinct infalling groups, or are they largely coherent streams of individual galaxies? And does the fraction of group galaxies and filament galaxies vary with the mass or dynamical state of the cluster? 
Fig. \ref{fig:fractions} shows the fractions of mock galaxies outside the cluster’s \R\ and inside $5\times$\R\ in groups and in filaments (i.e., galaxies with a distance to the skeleton of $D_{\rm{skel}} < 1 \hMpc$) as a function of cluster mass (Fig. \ref{fig:groups_mass}) and relaxedness (Fig. \ref{fig:groups_relaxedness}). Each point represents the fraction in one cluster, the bands indicate the means of the point distributions and corresponding 1$\sigma$ errors. 

While unrelaxed clusters have accreted large amounts of material (including through groups) in their recent history, they have also rapidly grown their \R\ as a consequence. The fraction of cluster mass in subhalos inside \R\ at present day is high, but we do not see evidence that the fraction of galaxies in filaments (closer than 1$\hMpc$) and groups outside \R\ and within 5\R\ of the cluster is higher in unrelaxed clusters (Fig. \ref{fig:groups_relaxedness}).
Independent of masses and dynamical status, approximately 10 per cent of all mock galaxies outside 1\R\ can be found in groups and roughly 45\% (30\%) of all mock galaxies outside 1\R\ are in filaments where filament thickness is defined as $D_{\rm{skel}} < 1 \hMpc$ ($D_{\rm{skel}} < 0.7 \hMpc$, see Sec. \ref{subsec:filament_identification}).
Most group host halos are located in filaments: 93\% of all group centres are located within 1Mpc/h of filament spines. Note that this reduces to 77\% for a more restricted filament thickness definition of 0.7Mpc/h. It is not surprising that most groups are part of filaments given \disperse\ identifies filaments by connecting maxima in the density field -- an a-postiori confirmation of the filament extraction. Nevertheless, this could be an important consideration for pre-processing studies since group galaxies in filaments have been shown to experience increased pre-processing compared to group galaxies outside filaments \citep{Poudel2017}. 

The low fraction of galaxies in groups may at first appear in tension with recent observational studies that typically report higher fractions \citep[e.g.,][]{McGee2009, dressler13, Cybulski2014}. We caution that a comparison is not straightforward given the differences in defining groups and mass thresholds. As discussed earlier, our cautiously identified group members represent galaxies that have spent a significant amount of time as part of groups and are thus likely to be environmentally effected by the group. \citet{Han2018} found that only $\sim$12\% of cluster members have spent more than 4 Gyr in a group and have therefore had enough time to quench \citep[satellite star formation rates evolve unaffected for 2 -- 4 Gyr after infall, ][]{wetzel13}. Many more galaxies spend only a limited amount of time (half of the galaxies spent less than 2.5 Gyr) in the host before joining the cluster population. Note also that observational analyses are complicated by high contamination rates, especially in the most typical groups that only host a few galaxies. It is important to keep in mind that observationally defined groups may include an additional 40\% of interlopers as group members \citep{Eke2004}.

Just like clusters, groups grow over time by merging and accreting members from their surroundings -- in most cases, this will be from the filament environment. We therefore investigate whether the fraction of group galaxies changes as a function of distance to the cluster centre. An increase could imply that even our cautious selection overestimates the fraction of group galaxies that have had enough time to be efficiently pre-processed, e.g., quenched as group satellites during infall. 
Fig. \ref{fig:frac_distance_all} shows the fraction of galaxies in filaments (solid line, black error band), in groups (dashed line, red error band) and backsplash galaxies (dotted line, yellow error band) as a function of distance to the cluster centre. Fractions are calculated in 30 shells out to 5\R\ surrounding the cluster. 
We do not show fractions inside \R, because at very small distances the volume of filaments quickly encompasses the entire volume, and fractions become meaningless. 
The red dot-dashed line in Fig. \ref{fig:frac_distance_all} shows that the fraction of galaxies in groups remains constant with distance. Similarly, we found that the average richness of groups stays constant as a function of distance. Richness is defined as the number of cluster members, i.e., all galaxies within \R\ of the group host. This is independent of whether they are located within filaments or outside of filaments. While this may seem in contrast to observations that report that groups in filaments have more satellites than outside of filaments \citep{Guo2015}, we again point towards differences (and difficulties) in defining groups consistently in simulations and observations and refer to our reasoning and choices (Sec. \ref{subsec:group_identification}). The constant fraction of galaxies in groups suggests that we indeed capture galaxies that have had a chance to pre-process. 

The black band (solid line) in Fig. \ref{fig:frac_distance_all} shows the \textit{fraction of all galaxies} in filaments as a function of distance, the grey line is the volume-corrected fraction.
Because the galaxy density and the relative volume of filaments increases towards clusters\footnote{Depending on the mass of the cluster, between 20 and 30\% of the volume immediately outside \R\ is taken up by filaments (calculated in a shell of 100 kpc thickness). }, the measured fraction of filament galaxies naturally increases. This can be seen by the upturn of the black solid line at smaller distances to the cluster. We reproduce and correct for this by calculating and subtracting the fraction of galaxies in randomly placed filament networks, which is shown by the dot-dashed line and light grey error band, i.e., for each cluster we calculate the fraction of galaxies in a network from another random cluster. The resulting volume-corrected fraction is shown in the solid dark grey line. For completion, we note that we have tested randomized orientations of the same cluster network as well as networks of a random different cluster for this correction. While results are not identical, both are valid ways to demonstrate the volume correction and differences are at the level of 10\% at small distances to the cluster centre.
The correction removes the increase of galaxies towards the cluster centre and flattens the curve -- a slight divergence from our results based on reference networks extracted from the underlying gas distribution discussed in \citet{Kuchner2020}.
In this "best case scenario" of gas filaments, we had found a small increase of galaxies in filaments closer to clusters (by about 8\%). However, given our choices for filament extraction, we found that it was most challenging to correctly identify filaments very close to clusters. As a consequence, such small effects may not have been picked up.

\subsubsection{Scatter on the extremes}
\label{subsec:scatter}

\threehundred simulations include 56 cluster volumes with very rich infalling groups of more than 150 members. These large groups can be treated as cluster-like systems with their own filament networks which will eventually merge with the more massive cluster. In \citet{Kuchner2020} we have shown that these second most massive halos (SMH) are connected to the central clusters with thick bridges, as has also been described in numerous observations \citep[e.g.,][]{Durret2008, Tanimura2019a, Umehata2019, Reiprich2021}. 

As a consequence, the "contamination" of filaments with group galaxies varies strongly across the sample (Fig. \ref{fig:scatter}) and we do not find a correlation of contamination with cluster properties. Overall, the contamination, i.e, the number of groups in filaments in cluster outskirts or their richness does not depend on the mass or dynamical state of the central cluster itself. The example figure shows fractions of galaxies in filaments and in groups in three clusters of the sample: the left panel shows a system where almost all filament galaxies are pristine filament galaxies. The contamination of filament galaxies that are in groups are shown in grey in the lower panel. We can identify two areas (at distance $\sim$2 and $\sim$5\R) with groups. At these distances, $\sim$20\% of the filament galaxies are in groups. The example in the middle shows one larger group embedded in a rich filament. The example on the right highlights a complex system with two large groups (akin to lower-mass clusters) that will merge with the cluster in the future: at two separate distances from the cluster centre, groups dominate the filaments and therefore around half of all filament galaxies in this system are found in groups.


\subsection{Backsplash galaxies in filaments}
\label{subsec:backsplash_in_filaments}
\begin{figure}
  \centering
  \subfloat[trajectory of backsplash galaxies in relaxed clusters]{\includegraphics[width=0.47\textwidth]{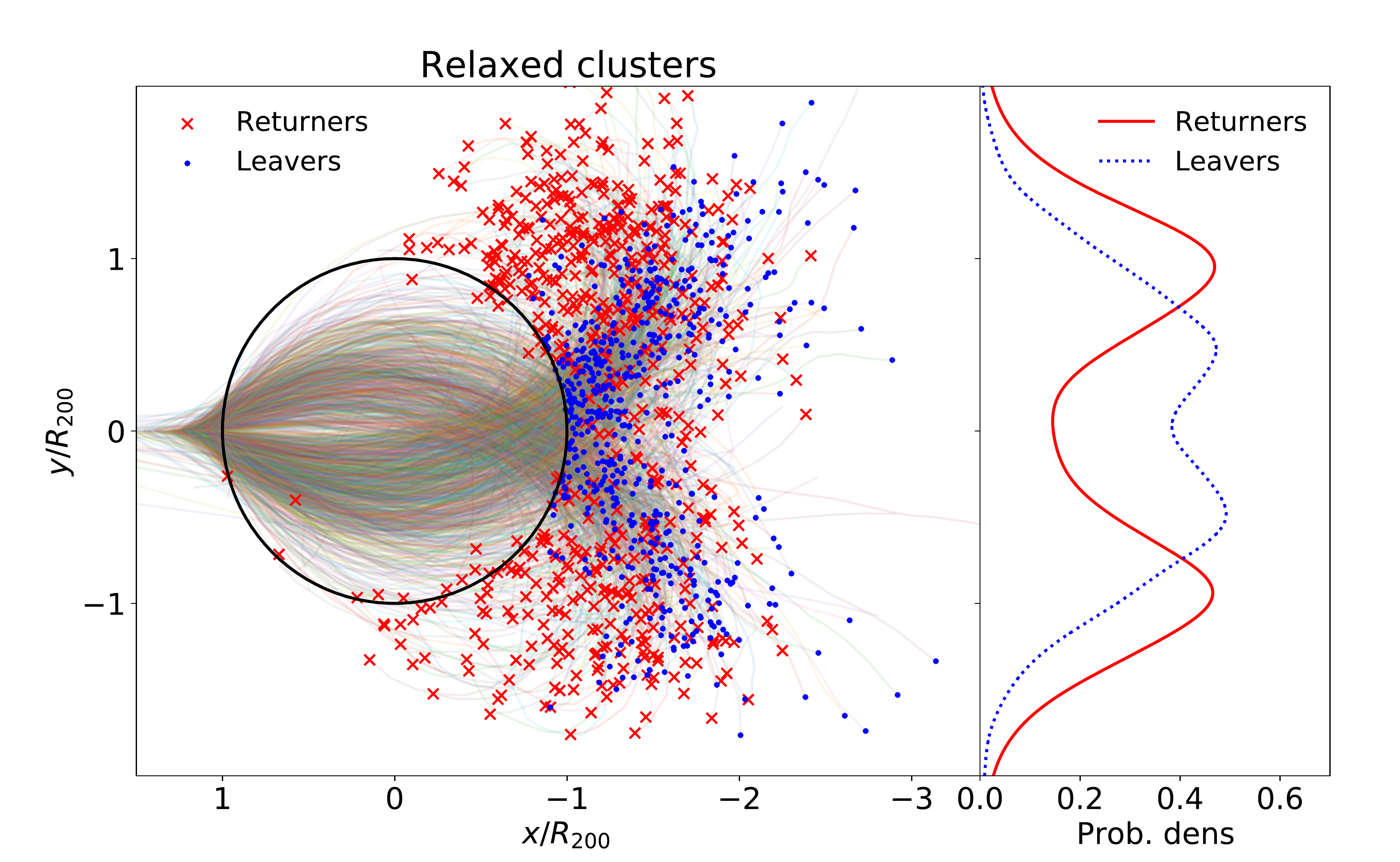}\label{fig:backsplash3}}
  \hfill
  \subfloat[trajectory of backsplash galaxies in unrelaxed clusters]{\includegraphics[width=0.47\textwidth]{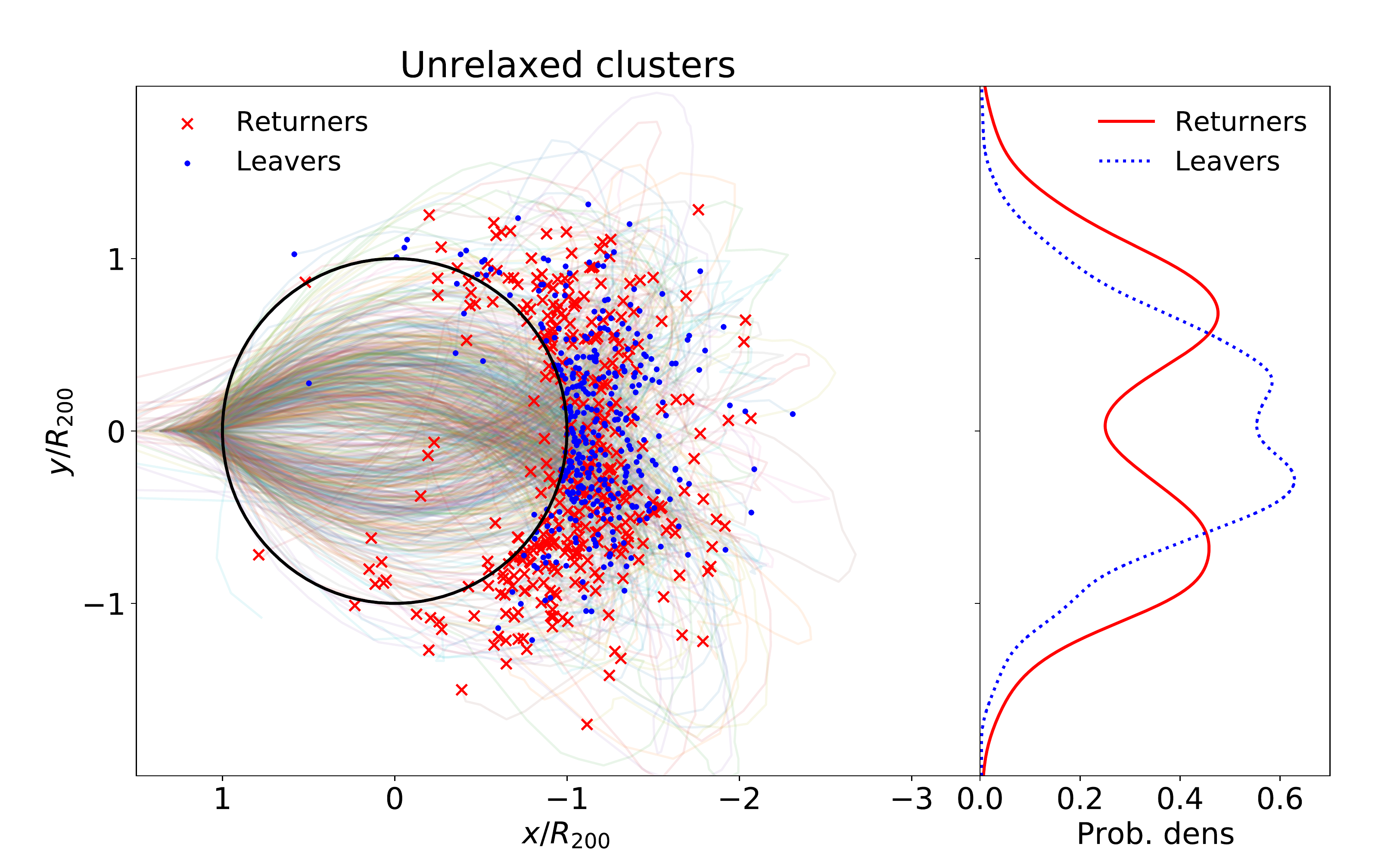}
  \label{fig:backsplash4}}
  \caption{Galaxies deflect from a straight line on their way through and out of the cluster. They leave and return to the cluster in a wide cone. Shown are traces of backsplash galaxies in relaxed (top) and unrelaxed (bottom) clusters at $z=0$.  "Leavers" are marked with blue dots and dotted line; they have not yet reached apocenter. "Returners", in red crosses and solid line, are on their next infall to the cluster. For clarity, points and tracks are shown for 3 clusters each, representative of the whole sample. The kernel density estimation on the right hand panel shows the degree of deviation from a straight line through the cluster for returners and leavers.}
  \label{fig:backsplash_investigation2}
\end{figure}
\begin{figure*}
  \centering
  \subfloat[fraction of galaxies and backsplash galaxies that are in filaments]{\includegraphics[width=0.47\textwidth]{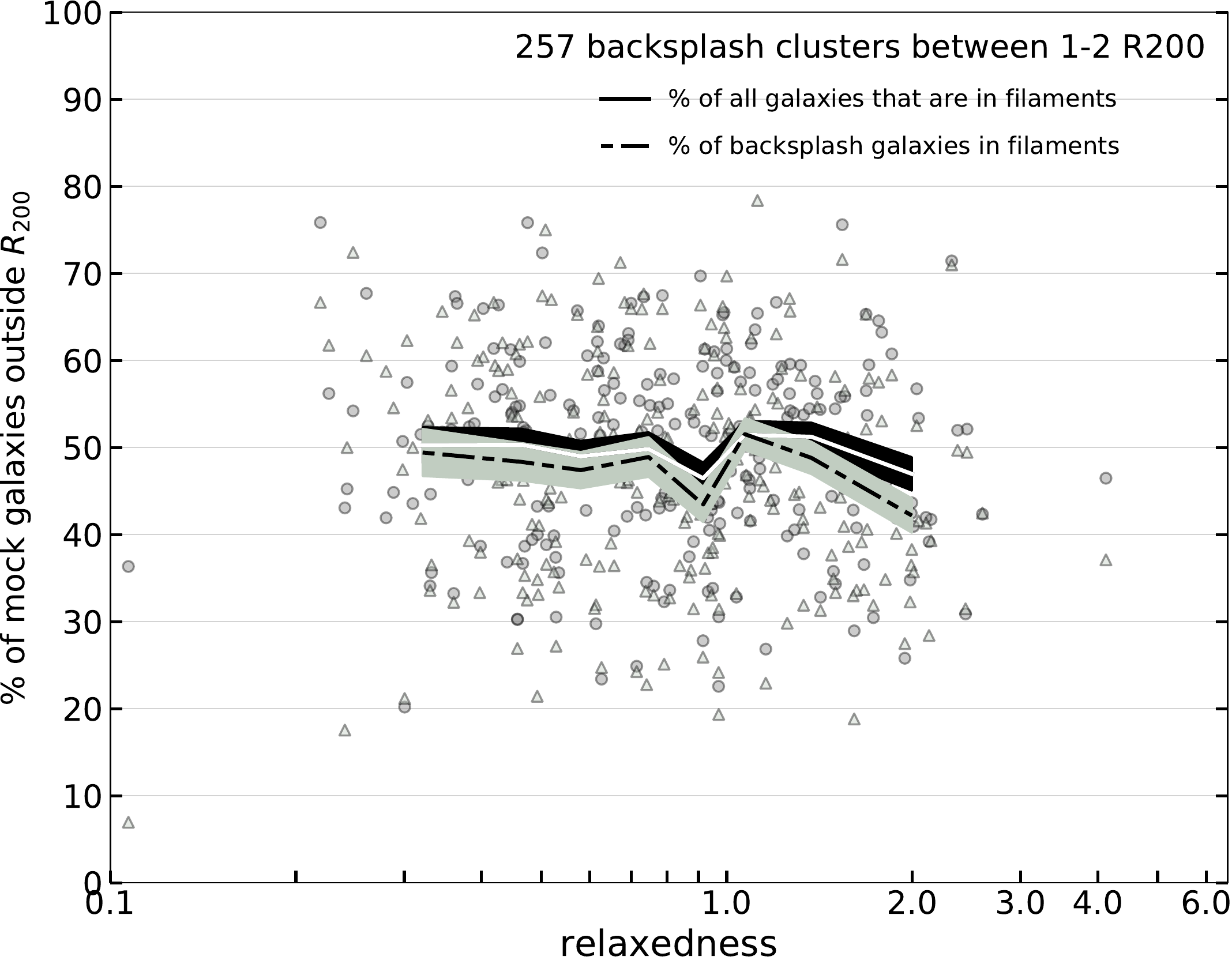} \label{fig:backsplash_a}}
  \subfloat[fraction of galaxies and filament galaxies that are backsplash]{\includegraphics[width=0.47\textwidth]{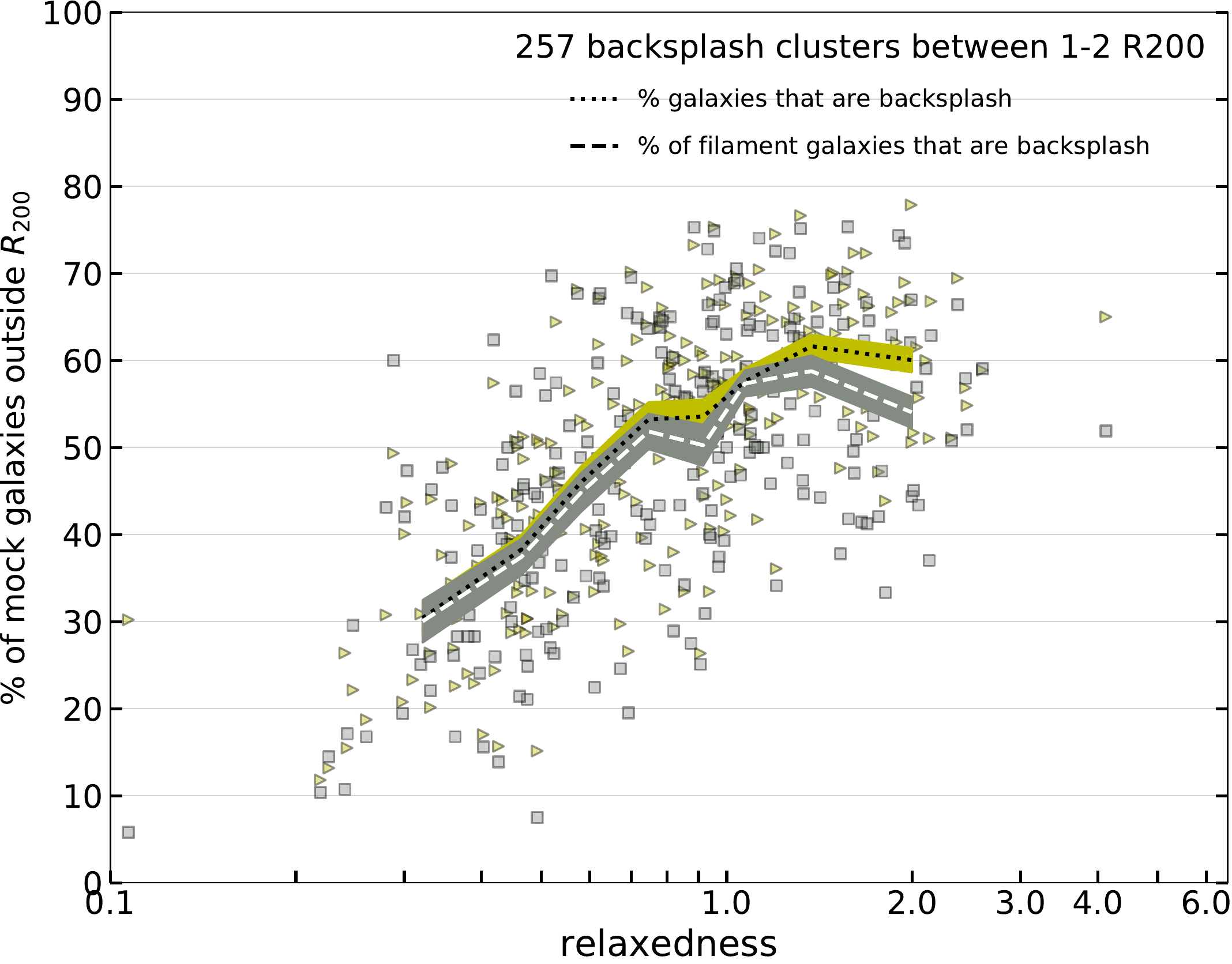}
  \label{fig:backsplash_b}}
  \caption{The fraction of backsplash galaxies in filaments of $1 \hMpc $ thickness is nearly identical to the fraction of all galaxies in filaments. This is independent of the dynamical state of the cluster (left, dot dashed line, grey error band). However, the overall fraction of backsplash galaxies increases from 30\% in unrelaxed (R<1) to 60\% in relaxed (R>1) clusters (right, dotted line, yellow error band). Consequently, the fraction of backsplash galaxies in filaments increases in nearly the same way. Bands show 1 $\sigma$ errors on the mean. Shown are fractions within $1 \hMpc$ of filament spines. Note that this figure only reports galaxies within 1 and 2 \R\ and a reduced sample of 257 clusters due to the requirement of continuous snapshot tracking to before $z=0.5$.}
  \label{fig:backsplash_investigation}
\end{figure*}

It is challenging to unambiguously identify individual backsplash galaxies in observations, i.e., galaxies whose orbital trajectories  have taken them through and out of the cluster after first or second infall. Usually, a variety of signatures need to add up: their gas morphologies could be altered due to ram pressure stripping \citep{Haynes1984, Abramson2011, Jaffe2015} and their stellar masses are lower due to tidal stripping \citep{Poggianti2017, Ramatsoku2019}. Further, stellar spectra might indicate post-starburst signatures \citep{Paccagnella2017, Kelkar2019}. In addition, backsplash galaxies show a stronger radial alignment than infalling galaxies \citep{Knebe2020}. But more commonly, backsplash galaxies are identified in phase-space diagrams through their positions and velocities. Statistically, they have recession velocities comparable to that of the cluster and are found in its immediate vicinity. However, we do not yet know how backsplash galaxies relate to filaments feeding clusters, i.e., if they have a preferential location with respect to filaments. Knowing whether they preferentially lie inside or outside of filaments could help to identify them. In addition, knowing how many backsplash galaxies are in filaments creates awareness that some observational signatures of galaxies in filaments (that possibly look like evidence of group- or filamentary pre-processing) may in fact be due to the galaxy's past environmental history of having gone through the cluster. 

Close to the cluster, backsplash galaxies become an increasingly important ingredient of the galaxy population mix, which can be appreciated by looking back to Fig. \ref{fig:frac_distance_all} where the backsplash population is denoted by the dotted line and yellow error band. The average fraction of backsplash galaxies rises to $\sim 65\%$ close to the cluster and is virtually absent outside of $\sim$ 2.5 \R. Note, however, that backsplash galaxies extend far beyond the typical virial radius of a cluster ($\sim 1.5$\R) and extend to the splashback radius\footnote{$R_{sp}$ is a physically motivated definition of the halo boundary where particles reach the apocenter of their first orbit; typically in the range [1,2.5]$R_{\rm{vir}}$ \citep{Diemer2014}.}, beyond which material is not expected to be virialised. 
We further investigate the positions and paths of backsplash galaxies in Fig. \ref{fig:backsplash_investigation2}. It shows the distribution of a representative sample of backsplash galaxies at redshift $z=0$ around clusters (indicated by the black circle), relative to the position at which they first entered the cluster, and their trajectories, for relaxed (top) and unrelaxed (bottom) clusters separately. We produced this plot by rotating the path taken through the cluster by each backsplash galaxy, so that each galaxy is on the x-axis (y=z=0) in their last snapshot before entering the cluster. We also rotated the paths such that the motion in the z-direction is minimised, and hence the galaxy paths are (approximately) in the plane of the page. Backsplash galaxies leave a cluster typically after $\sim$2 Gyr opposite the location where they entered and build a "cone" of opening angle $23^{+14}_{-12}$ degrees in relaxed clusters and s$21^{+17}_{-11}$ degrees in unrelaxed clusters. They return in a wider cone of $41^{+20}_{-16}$ degrees in relaxed clusters and $35^{+22}_{-16}$ degrees in unrelaxed clusters \citep[see also][]{Knebe2004}. The angles are the median values and 1-sigma spread, where 0 degrees corresponds to a galaxy that has passed straight through a cluster with no deviation (y=0 in Fig. \ref{fig:backsplash_investigation2}). For both relaxed and unrelaxed clusters, the returners are slightly more deflected. 

Following this picture, it is easy to imagine that if a galaxy falls in through a filament, chances are high that that the backsplash galaxy will collapse onto a filament on the other side of the cluster and thus return as part of filaments. This is because filaments are not randomly positioned either: they 
preferentially follow the semi-major axis of the main halo or connect to their second most massive halo as a bridge \citep{Kuchner2020}. 
In addition, so far we cannot rule out that backsplash galaxies (help to) \textit{form} a filament when they return to the cluster. To investigate this, we divide backsplash galaxies by their velocities into leavers and returners, i.e, galaxies that have gone through the cluster and are moving away from it in the final snapshot (either for the first or second time) are labelled as "leavers" (blue dots and dotted line in Fig. \ref{fig:backsplash_investigation2}) and galaxies that have gone out, turned around and are approaching the cluster (either for the first or second time) are labelled "returners" (red 'x' and solid line). 
In other words, "leavers" have left the cluster but have not yet reached the apocentre of their orbit, while "returners" have passed apocentre, and are now on a second or further infall towards the cluster. Note that all 257 clusters were used in this analysis (see Sec. \ref{subsec:backsplash_identification}), however for clarity we only show the paths and final positions of galaxies in three clusters, which are representative of the larger sample. 

Galaxies that have gone through the cluster are likely to have been deflected from the central axis they each start from, rather than pass straight through. This can be inferred from the double-peaked probability distributions of the y-coordinate of galaxy positions at redshift $z=0$ in the right panels of Fig. \ref{fig:backsplash_investigation2}. It represents the distance each galaxy has deviated from a straight path through the cluster and can be seen in both relaxed and unrelaxed clusters\footnote{The kernel density estimation (KDE) is made using data from all clusters, not just those shown in the left panel of the plot.}. Following the expectation that clusters are embedded in a large cosmic filament extended along the major axis, we can now anticipate that backsplash galaxies deviate from a major filament. This is supported by our finding in \citet{Rost2020a} where we found that gas preferentially falls into nodes inside filaments, but preferentially leaves the cluster outside filaments. 
Returners are more heavily deflected sideways, in both relaxed and unrelaxed clusters. 

In this analysis, every backsplash galaxy is constructed to start at the same point. In reality, backsplash galaxies enter the cluster from a number of positions around the cluster --  through filaments, as groups and as isolated galaxies -- and each one deflects and scatters dynamically.  The many infall and therefore scatter directions add up to create a cloud of backsplash galaxies, which can be appreciated as yellow points around the cluster in Fig. \ref{fig:groups}. Importantly, this cloud of additional galaxies close to \R\ does not influence  the filament finding process. In practice, the homogeneous cloud of backsplash galaxies close to the cluster is not an important feature for \disperse, provided enough volume or area is available.
Even in a (hypothetical) extreme case where all galaxies come in through filaments (and we know from Fig. \ref{fig:fractions} that statistically this is not the case), they leave the cluster scattered in wide cones that overlap, again smearing out to a cloud of backsplash galaxies that is very similar to the overall distribution of infalling galaxies. We thus see very little evidence that backsplash galaxies are distributed differently to infalling galaxies with respect to filaments. 
This is evident in Fig. \ref{fig:backsplash_a}, which shows the percentage of all galaxies in filaments (black points, solid line, black error band) and that of backsplash galaxies in filaments (grey triangles for individual points and dot-dashed line, grey error band for the 1$\sigma$ error on the mean). The two curves are nearly identical, signifying that backsplash galaxies are neither more nor less likely to re-enter the cluster through filaments than a galaxy on its first infall. We therefore see no evidence that backsplash galaxies collapse to form a filament. Filaments are stable geometrical features that do not quickly change or form. 
However, Fig. \ref{fig:backsplash_investigation2} suggests that there may be a difference between galaxies leaving and returning in relaxed and unrelaxed clusters.

\subsubsection{Dependence on dynamical state of the cluster}

The fraction and extent of backsplash galaxies around clusters not only varies strongly with distance to the cluster but also with dynamical state of the cluster. Fig. \ref{fig:backsplash_investigation2} shows that backsplash galaxies around relaxed clusters spray further than in unrelaxed clusters, where the entire backsplash population is typically contained within 2\R. 
To investigate whether this resulted from the fact that our relaxed clusters have a lower average radius, we reproduced these plots, normalising by 2 Mpc (which is approximately the average cluster radius) instead of \R. These plots are not shown, but changing this normalisation had very little effect on the results. Rather, this difference is due to the rapid increase of the cluster's radius following mergers that lead to unrelaxed dynamical states -- faster than backsplash galaxies replenish  \citep{Haggar2020}.
The difference is significant: 
The fraction of backsplash galaxies increases from 30\% in unrelaxed (R<1) to 60\% in relaxed (R>1) clusters (dashed line and yellow error band in Fig. \ref{fig:backsplash_b}). As a direct consequence of Fig. \ref{fig:backsplash_a}, the fraction of backsplash galaxies in filaments rises at the same rate (dashed line, grey error band).
We see some hints of a deviation in the most relaxed clusters of the sample, in the sense of a lower fraction of backsplash compared to infalling galaxies in filaments.  
\begin{figure}
	\includegraphics[width=\columnwidth]{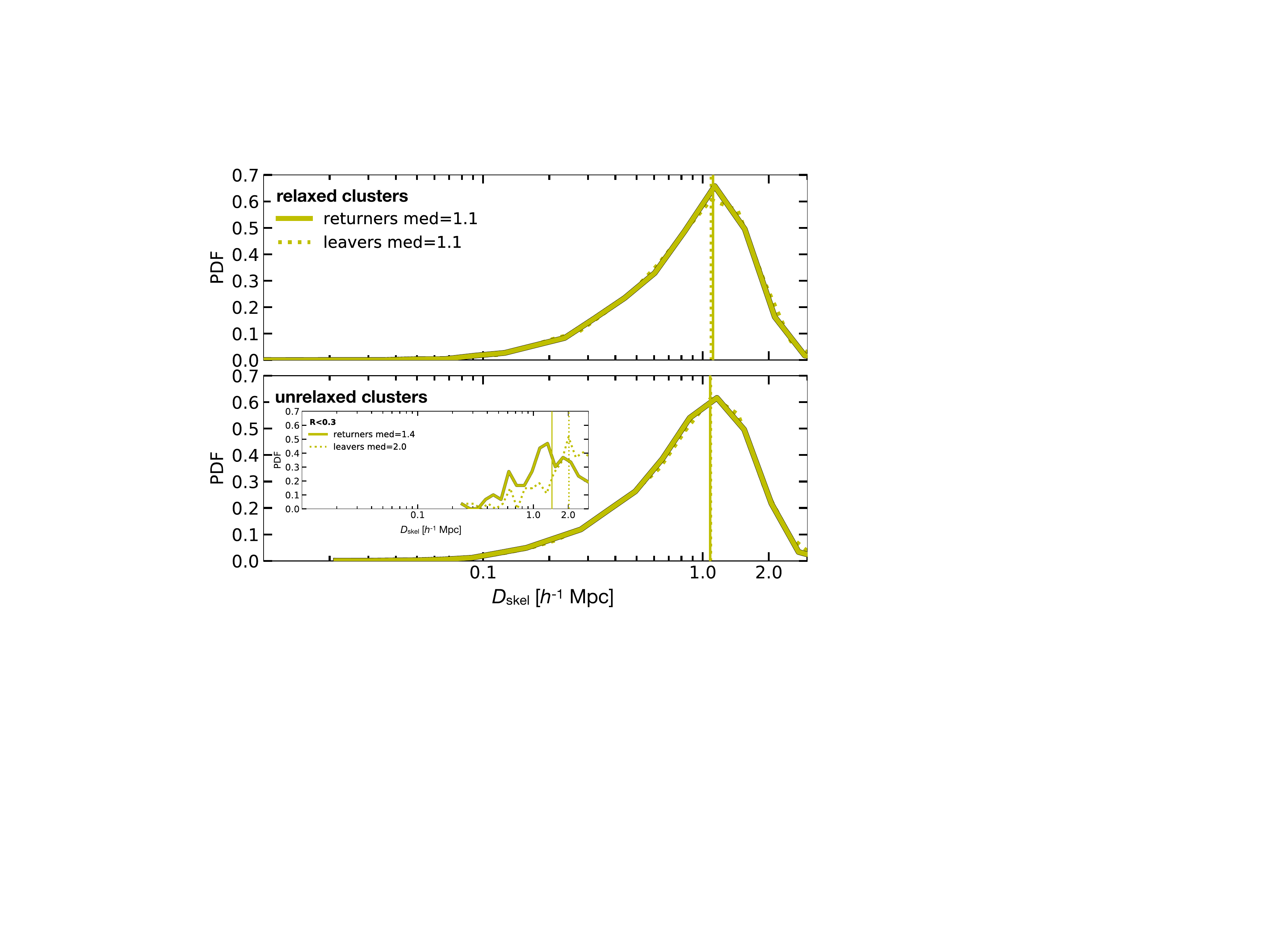}
    \caption{Despite the different distributions of backsplash galaxies in relaxed and unrelaxed clusters, they are distributed in the same way in filaments around relaxed and unrelaxed clusters. Because they are distributed close to the R200, a lot of the volume is actually made up of filaments. The yellow curve and the dot-dashed curve rise in similar ways at least inside 1.5R200, where backsplash galaxies are dominant. }
    \label{fig:leavers_remainers}
\end{figure}

Leavers and returners may be more clearly separable in relaxed clusters than in unrelaxed clusters (compare the two peaks in the KDE of the right panel Fig. \ref{fig:backsplash_investigation2}), but we do not see a dependence on dynamical state of the cluster in relation to filaments: backsplash galaxies are distributed in the same way with respect to filaments, whether they are in relaxed or unrelaxed clusters and whether they are leaving or returning to the cluster (explained by the picture of a homogenous cloud of backsplash galaxies due to the scatter dynamics of galaxies passing through the cluster, as discussed in the previous section).
Fig. \ref{fig:leavers_remainers} underpins this uniformity. It shows PDFs of measured distances from leavers and returners to filament spines in relaxed (top) and unrelaxed (bottom) clusters. Clearly, there is no difference between returners and leavers and also no difference in relaxed and unrelaxed clusters, with the small exception of very unrelaxed clusters (insert in Fig. \ref{fig:leavers_remainers}). In a sample of the most unrelaxed clusters with relaxedness parameters R<0.3, returners came back to the clusters significantly closer to filament spines than they had left the cluster. This may indicate some memory of a merger, where in unrelaxed clusters the preferential direction of velocities before the merger is retained, and in relaxed clusters this axis was lost when velocities randomised. 
For the majority of clusters, however, we see no dependence of the location of backsplash galaxies in relaxed and unrelaxed and in leavers and returners with respect to filaments. 
Note that the area where backsplash galaxies prevail is a turbulent region characterized by accretion shocks where the infalling gas is significantly slowed down and heated while becoming part of the intracluster medium. The turbulence close to the cluster induced by the mixing of material that collapses towards filaments, as well as into the cluster, and gas shocks triggered by substructures is further complicated by signatures of backsplash galaxies. We described these complex gas velocity fields close to \threehundred clusters in \citet{Rost2020a}.

Finally, some backsplash galaxies may be in groups if they have fallen in as part of groups and leave \R\ still maintaining group membership. Following the trajectories of groups in \threehundred, we found that backsplash groups are relatively rare since infalling groups lose the majority of their members inside \R\ of the cluster. With our group definition, only 9\% of backsplash galaxies are members of a group at cluster infall.
 



\section{Conclusions: Heterogeneous filament environments}

\begin{figure}
	\includegraphics[width=\columnwidth]{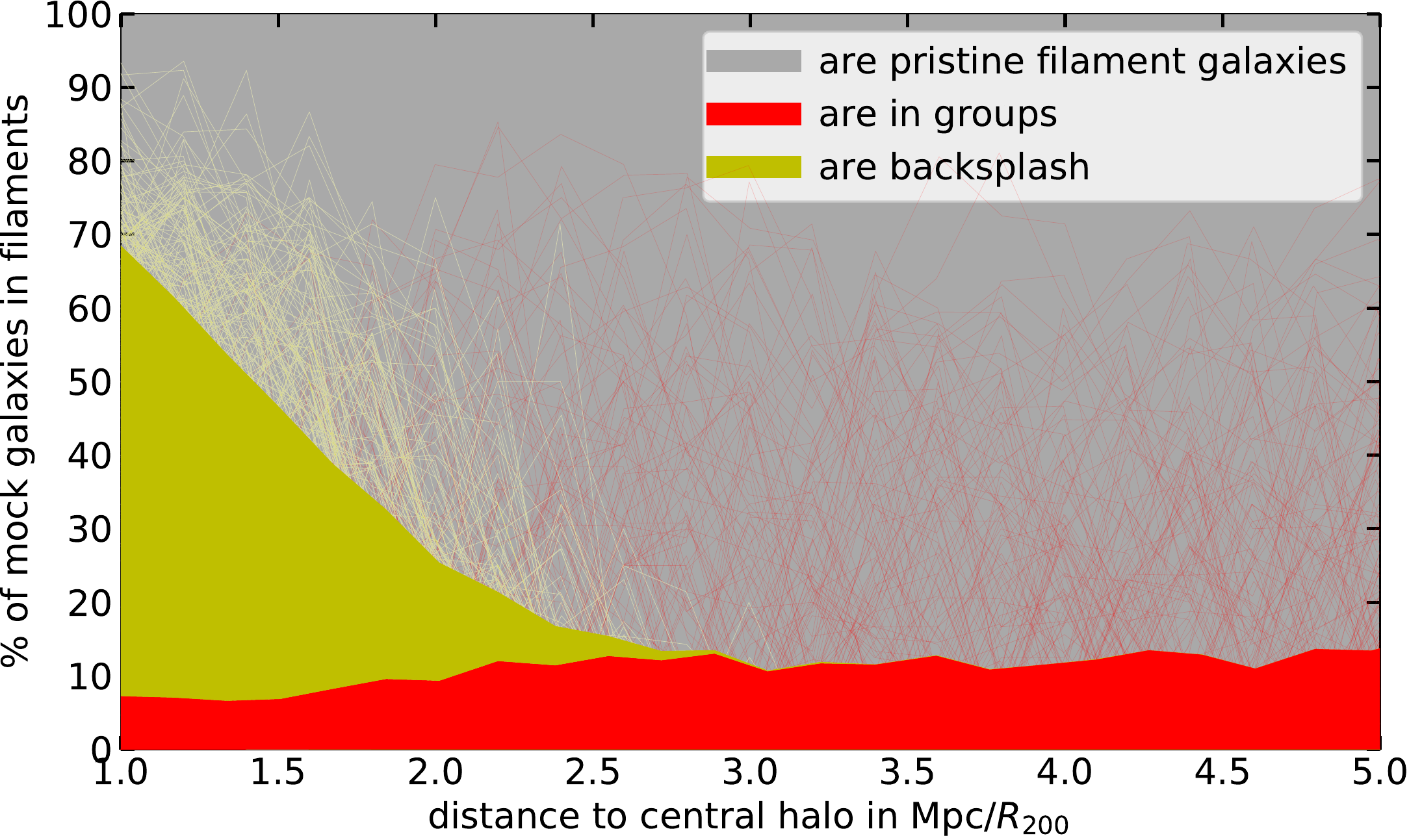}
    \caption{As a way to compare total values of filament members across distances from the cluster (\R=1.0), we plot a stacked histogram. It shows the expected fractions of galaxies in each environment within filaments of constant thickness as a function of distance to the cluster centre normalized by \R: group galaxies as defined in Sec. \ref{subsec:group_identification} (red area), backsplash galaxies (yellow area) and "pristine" filament galaxies (grey area). By inference, only about 30\% of all galaxies that fall into the cluster through filaments are "pristine". The thin lines highlight the large cluster-by-cluster variations. }
    \label{fig:inventory}
\end{figure}

Cosmic filaments that feed clusters host galaxies with diverse star formation histories. 
The galaxies may get affected by their current environment, either denser large-scale filaments and/or groups that are part of the filament network ("pre-processed"). This could lead to measurable changes, e.g., of the gas content, star-formation activity and galaxy morphology. Others may have been processed in the past by the galaxy cluster during their first infall. The galaxy mixture strongly depends on the distance from the cluster core and dynamical state of the cluster. Understanding the constituents of galaxy cluster outskirts as a combination of different environments, where the important environment of filaments themselves are heterogeneous, helps to better understand the nature and relative importance of environmental processes on galaxy mass assembly and quenching.
Fig. \ref{fig:inventory} summarises this non-uniform environment and shows an inventory of galaxies in filaments around simulated \threehundred\ clusters, a benchmark to compare observational signatures with. These numbers are based on a characteristic filament core thickness of $1\hMpc$ and halo masses of $M_{\rm{halo}} > 3\times 10^{10} \hMsun$ (comparable to $M_{\rm{*}} > 3\times 10^{9} \hMsun$). They change according to choices that will depend on the individual science case and emphasis on e.g., purity, completeness, accuracy or precision\footnote{We refer the reader to \citet{Kuchner2020} for a detailed overview of how these choices may bias expectations.}. Importantly, fractions do not depend on cluster halo mass. The figure summarises the composition of filaments feeding clusters as a function of distance to the cluster centre. From it we conclude the following: 
\begin{figure}
	\includegraphics[width=\columnwidth]{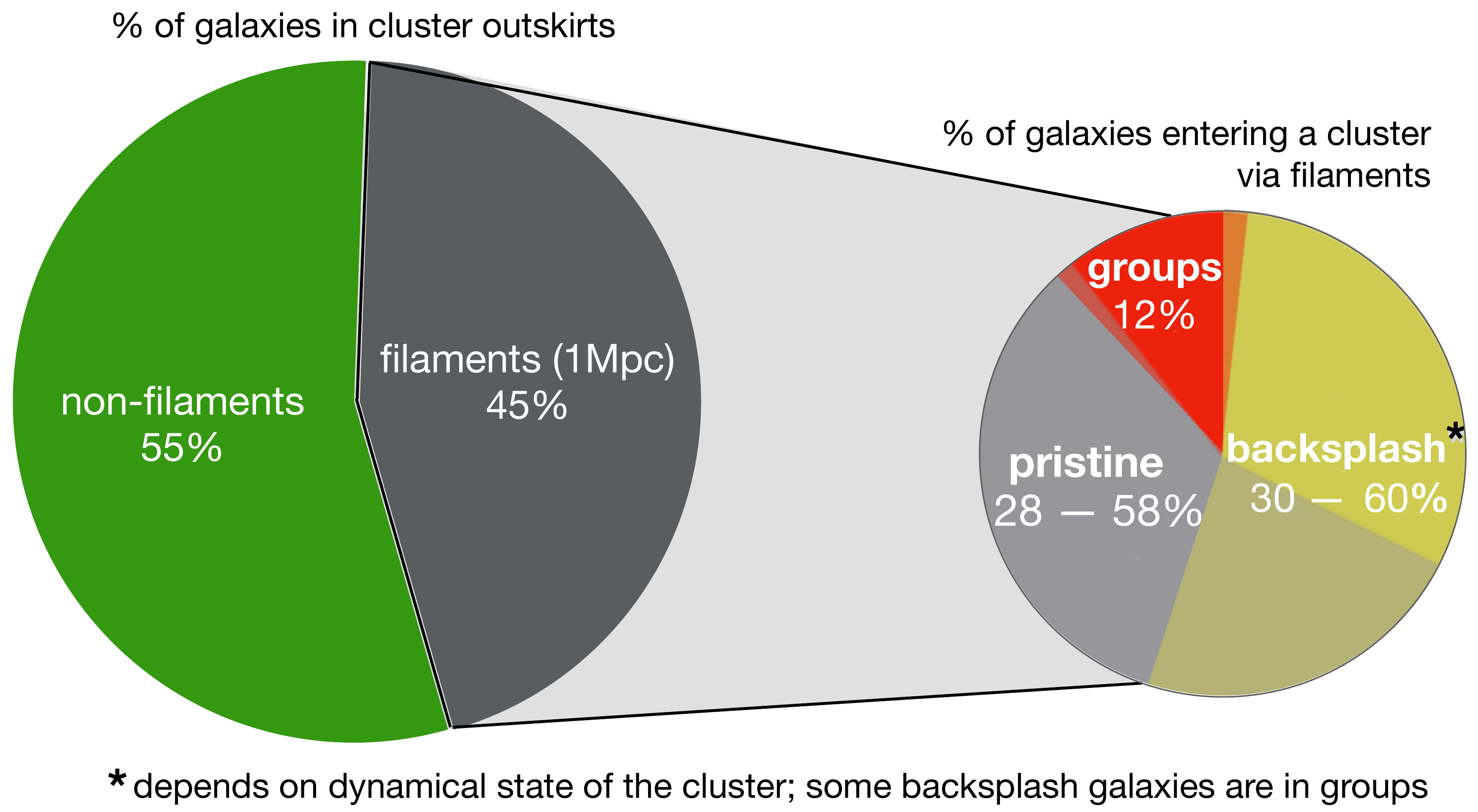}
    \caption{
    Up to 45 per cent of galaxies accreted by clusters are closer than 1$ \hMpc$ to a filament spine, which we define as being "inside filaments" (see text for a discussion on choosing an optimal filament thickness). Filaments themselves are heterogeneous environments that host groups, and backsplash galaxies alongside galaxies that have been environmentally effected by the cosmic filament alone. The pie chart on the right details the breakdown of galaxies in different environments inside filaments at \R. The number of backsplash galaxies is highly dependant on the cluster's dynamical state.}
    \label{fig:summary}
\end{figure}

\begin{itemize}
\item{\textbf{Group galaxies:} 12 percent of all filament galaxies in cluster outskirts (between 1 and 5 \R\ of the cluster) are located in groups in filaments, where we expect pre-processing by group environments. This number is highly dependent on the exact definition of group membership (Sec. \ref{subsec:group_identification}). In the context of \threehundred and keeping observational challenges and goals of pre-processing studies in mind, we define groups as galaxies within 1\R\ of a halo with $\sigma_v > 300\ h^{-1}\rm{km/s}$. This likely captures the correct number of galaxies that has spent a significant time (longer than 4 Gyrs) in groups \citep{Han2018}. The fraction of galaxies in groups doubles when this criterion is lowered to 150$h^{-1}\rm{km/s}$. 90\% of group hosts are located in filaments, owing to a large extend to the fact that they mark maxima in the galaxy distribution which are used to construct the filament network (see Sec. \ref{subsec:filament_identification}). 
While there is considerable cluster-to-cluster variation (Fig. \ref{fig:scatter}), on average the fraction of group galaxies in filaments remains constant with distance from the cluster.}
\item{\textbf{Backsplash galaxies:} close to the cluster centre, between 30 percent (in unrelaxed) and 60 percent (in relaxed clusters) of all galaxies are members of the backsplash populations (Fig. \ref{fig:backsplash_b}), i.e., they have been processed by the cluster. This number is highly dependent on the dynamical state of the cluster and distance to the cluster centre: we find more backsplash galaxies in relaxed clusters and close to \R. The number drops sharply with increasing distance and we find no backsplash galaxies beyond 2.5\R. The increasing prevalence of backsplash galaxies around clusters make it challenging to disentangle the post-processing effects of clusters and the pre-processing of cosmic web environments. Backsplash galaxies are deflected on their pass through the cluster and scatter on their way out which produces a cloud of backsplash galaxies around the cluster. Therefore, they have no preferred location with respect to filaments, i.e., they are not more likely to fall back onto clusters through filaments (Fig. \ref{fig:backsplash_a}) -- neither in relaxed nor unrelaxed clusters (Fig. \ref{fig:leavers_remainers}).  
}
\item{\textbf{Pristine filament galaxies}: The remaining $\sim 33\%$ of galaxies in filaments at cluster \R\ are "pristine filament galaxies". These are galaxies entering a cluster via coherent streams of individual galaxies. Importantly, this scenario assumes filaments of constant thickness -- a simplification, since filaments are likely growing thicker closer to massive nodes with an increase of galaxies in filaments (see \citet{Kuchner2020} for a discussion). }
\end{itemize}

The complex cluster outskirt physics make the reconstruction of environmental histories of galaxies falling into clusters not only challenging but dependent on factors such as the dynamical state of the cluster or the distance to the cluster centre. Measurements are challenging since this is a regime where the infall, merging and virialisation of matter intertwine. Near clusters, accretion shocks and backsplash galaxies dominate and complicate the velocities of galaxies and measurements of their host environments. Further out, galaxy groups and large-scale filaments of the cosmic web may take over. Each relate to specific environmental mechanisms and thus influence expectations for observational evidence of pre-processing (observed effects due to increased densities) in galaxies around clusters. The results presented in this paper demonstrate a statistical breakdown of galaxies in cluster outskirt environments, emphasising the variety of environments and environmental histories galaxies in filaments can have and typical journeys of galaxies before falling into clusters. Groups and filaments are the instantaneous environment we find galaxies in, backsplash galaxies contain a record of where the galaxies have been in the past. In addition, these are not absolutes: some backsplash galaxies are in filaments, some are in groups, some are in the remaining "field" around the cluster. In summary, while up to 45\% of all galaxies fall into clusters via filaments (closer than $1\hMpc$ from the extracted filament spine), filaments themselves are heterogeneous environments that host groups and backsplash galaxies, alongside a minority of galaxies that have been environmentally effected by the cosmic filament alone (Fig. \ref{fig:summary}).

\section*{Acknowledgements}

We thank the referee for providing useful feedback to this study. 
This work has been made possible by \threehundred collaboration, which benefits from financial support of the European Union’s Horizon 2020 Research and Innovation programme under the Marie Sk\l{}odowskaw-Curie grant agreement number 734374, i.e. the LACEGAL project. \threehundred simulations used in this paper have been performed in the MareNostrum Supercomputer at the Barcelona Supercomputing Center, thanks to CPU time granted by the Red Espa\~{n}ola de Supercomputaci\'{o}n.
UK acknowledges support from the Science and Technology Facilities Council through grant number RA27PN. 
AK is supported by the Ministerio de Ciencia, Innovaci\'{o}n y Universidades (MICIU/FEDER) under research grant PGC2018-094975-C21 and further thanks Matt Haynes and Clare Wadd for Sarah Records.
WC is supported by the European Research Council under grant number 670193 and by the STFC AGP Grant ST/V000594/1. He further acknowledges the science research grants from the China Manned Space Project with NO. CMS-CSST-2021-A01 and CMS-CSST-2021-B01.
The authors contributed to this paper in the following ways: UK, AAS, MEG and FRP formed the core team. UK identified filaments, analysed the data, produced the plots (with the exception of Fig. 5) and wrote the paper with ongoing input from the core team and comments from co-authors. RH calculated the relaxedness parameter $R$ based on dynamical state parameters produced by WC and provided identification of the backsplash galaxies incl. Fig. 5; GY supplied the simulation data; AK the halo catalogues. 

\section*{Data Availability}
Data available on request to \threehundred collaboration: https://www.the300-project.org.




\bibliographystyle{mnras}
\bibliography{references.bib} 

\begin{thebibliography}{}
\makeatletter
\relax
\def\mn@urlcharsother{\let\do\@makeother \do\$\do\&\do\#\do\^\do\_\do\%\do\~}
\def\mn@doi{\begingroup\mn@urlcharsother \@ifnextchar [ {\mn@doi@}
  {\mn@doi@[]}}
\def\mn@doi@[#1]#2{\def\@tempa{#1}\ifx\@tempa\@empty \href
  {http://dx.doi.org/#2} {doi:#2}\else \href {http://dx.doi.org/#2} {#1}\fi
  \endgroup}
\def\mn@eprint#1#2{\mn@eprint@#1:#2::\@nil}
\def\mn@eprint@arXiv#1{\href {http://arxiv.org/abs/#1} {{\tt arXiv:#1}}}
\def\mn@eprint@dblp#1{\href {http://dblp.uni-trier.de/rec/bibtex/#1.xml}
  {dblp:#1}}
\def\mn@eprint@#1:#2:#3:#4\@nil{\def\@tempa {#1}\def\@tempb {#2}\def\@tempc
  {#3}\ifx \@tempc \@empty \let \@tempc \@tempb \let \@tempb \@tempa \fi \ifx
  \@tempb \@empty \def\@tempb {arXiv}\fi \@ifundefined
  {mn@eprint@\@tempb}{\@tempb:\@tempc}{\expandafter \expandafter \csname
  mn@eprint@\@tempb\endcsname \expandafter{\@tempc}}}

\bibitem[\protect\citeauthoryear{Abramson, Kenney, Crowl, Chung, van Gorkom,
  Vollmer  \& Schiminovich}{Abramson et~al.}{2011}]{Abramson2011}
Abramson A.,  Kenney J. D.~P.,  Crowl H.~H.,  Chung A.,  van Gorkom J.~H.,
  Vollmer B.,   Schiminovich D.,  2011, \mn@doi [The Astronomical Journal]
  {10.1088/0004-6256/141/5/164}, 141, 164

\bibitem[\protect\citeauthoryear{Ade et~al.,}{Ade et~al.}{2016}]{Ade2016}
Ade P. A.~R.,  et~al., 2016, \mn@doi [Astronomy {\&} Astrophysics]
  {10.1051/0004-6361/201525830}, 594, A13

\bibitem[\protect\citeauthoryear{Alpaslan et~al.,}{Alpaslan
  et~al.}{2016}]{Alpaslan2016}
Alpaslan M.,  et~al., 2016, \mn@doi [Monthly Notices of the Royal Astronomical
  Society] {10.1093/mnras/stw134}, 457, 2287

\bibitem[\protect\citeauthoryear{Arzoumanian et~al.,}{Arzoumanian
  et~al.}{2019}]{Arzoumanian2019}
Arzoumanian D.,  et~al., 2019, \mn@doi [Astronomy {\&} Astrophysics]
  {10.1051/0004-6361/201832725}, 621, A42

\bibitem[\protect\citeauthoryear{Bah{\'{e}} \& McCarthy}{Bah{\'{e}} \&
  McCarthy}{2014}]{Bahe2014}
Bah{\'{e}} Y.~M.,  McCarthy I.~G.,  2014, \mn@doi [Monthly Notices of the Royal
  Astronomical Society] {10.1093/mnras/stu2293}, 447, 969

\bibitem[\protect\citeauthoryear{Bah{\'{e}}, McCarthy, Balogh  \&
  Font}{Bah{\'{e}} et~al.}{2013}]{Bahe2013}
Bah{\'{e}} Y.~M.,  McCarthy I.~G.,  Balogh M.~L.,   Font A.~S.,  2013, \mn@doi
  [Monthly Notices of the Royal Astronomical Society] {10.1093/mnras/stt109},
  430, 3017

\bibitem[\protect\citeauthoryear{Balcells et~al.,}{Balcells
  et~al.}{2010}]{Balcells2010}
Balcells M.,  et~al., 2010, in McLean I.~S.,  Ramsay S.~K.,   Takami H.,  eds,
  Ground-based and Airborne Instrumentation for Astronomy {III}. {SPIE},
  \mn@doi{10.1117/12.856947}

\bibitem[\protect\citeauthoryear{Balogh, Morris, Yee, Carlberg  \&
  Ellingson}{Balogh et~al.}{1997}]{Balogh1997}
Balogh M.~L.,  Morris S.~L.,  Yee H. K.~C.,  Carlberg R.~G.,   Ellingson E.,
  1997, \mn@doi [The Astrophysical Journal] {10.1086/310927}, 488, L75

\bibitem[\protect\citeauthoryear{{Balogh}, {Navarro}  \& {Morris}}{{Balogh}
  et~al.}{2000}]{balogh00}
{Balogh} M.~L.,  {Navarro} J.~F.,   {Morris} S.~L.,  2000, \mn@doi [\apj]
  {10.1086/309323}, \href {http://adsabs.harvard.edu/abs/2000ApJ...540..113B}
  {540, 113}

\bibitem[\protect\citeauthoryear{Beck et~al.,}{Beck et~al.}{2015}]{Beck2015}
Beck A.~M.,  et~al., 2015, \mn@doi [Monthly Notices of the Royal Astronomical
  Society] {10.1093/mnras/stv2443}, 455, 2110

\bibitem[\protect\citeauthoryear{Benavides, Sales  \& Abadi}{Benavides
  et~al.}{2020}]{Benavides2020}
Benavides J.~A.,  Sales L.~V.,   Abadi M.~G.,  2020, \mn@doi [Monthly Notices
  of the Royal Astronomical Society] {10.1093/mnras/staa2636}, 498, 3852

\bibitem[\protect\citeauthoryear{Ben{\'{\i}}tez-Llambay, Navarro, Abadi,
  Gottlöber, Yepes, Hoffman  \& Steinmetz}{Ben{\'{\i}}tez-Llambay
  et~al.}{2013}]{Benitez-Llambay2013}
Ben{\'{\i}}tez-Llambay A.,  Navarro J.~F.,  Abadi M.~G.,  Gottlöber S.,  Yepes
  G.,  Hoffman Y.,   Steinmetz M.,  2013, \mn@doi [The Astrophysical Journal]
  {10.1088/2041-8205/763/2/l41}, 763, L41

\bibitem[\protect\citeauthoryear{Berrier, Stewart, Bullock, Purcell, Barton  \&
  Wechsler}{Berrier et~al.}{2008}]{Berrier2008}
Berrier J.~C.,  Stewart K.~R.,  Bullock J.~S.,  Purcell C.~W.,  Barton E.~J.,
  Wechsler R.~H.,  2008, \mn@doi [The Astrophysical Journal]
  {10.1088/0004-637x/690/2/1292}, 690, 1292

\bibitem[\protect\citeauthoryear{Bianconi, Smith, Haines, McGee, Finoguenov  \&
  Egami}{Bianconi et~al.}{2017}]{Bianconi2017}
Bianconi M.,  Smith G.~P.,  Haines C.~P.,  McGee S.~L.,  Finoguenov A.,   Egami
  E.,  2017, \mn@doi [Monthly Notices of the Royal Astronomical Society:
  Letters] {10.1093/mnrasl/slx167}, 473, L79

\bibitem[\protect\citeauthoryear{Blanton \& Moustakas}{Blanton \&
  Moustakas}{2009}]{Blanton2009}
Blanton M.~R.,  Moustakas J.,  2009, \mn@doi [Annual Review of Astronomy and
  Astrophysics] {10.1146/annurev-astro-082708-101734}, 47, 159

\bibitem[\protect\citeauthoryear{Cautun, van~de Weygaert, Jones  \&
  Frenk}{Cautun et~al.}{2014}]{Cautun2014}
Cautun M.,  van~de Weygaert R.,  Jones B. J.~T.,   Frenk C.~S.,  2014, \mn@doi
  [Proceedings of the International Astronomical Union]
  {10.1017/s1743921316009613}, 11, 47

\bibitem[\protect\citeauthoryear{Colberg, Krughoff  \& Connolly}{Colberg
  et~al.}{2005}]{Colberg2005}
Colberg J.~M.,  Krughoff K.~S.,   Connolly A.~J.,  2005, \mn@doi [Monthly
  Notices of the Royal Astronomical Society]
  {10.1111/j.1365-2966.2005.08897.x}, 359, 272

\bibitem[\protect\citeauthoryear{Cui, Knebe, Yepes, Yang, Borgani, Kang, Power
  \& Staveley-Smith}{Cui et~al.}{2017}]{Cui2017}
Cui W.,  Knebe A.,  Yepes G.,  Yang X.,  Borgani S.,  Kang X.,  Power C.,
  Staveley-Smith L.,  2017, \mn@doi [Monthly Notices of the Royal Astronomical
  Society] {10.1093/mnras/stx2323}, 473, 68

\bibitem[\protect\citeauthoryear{Cui et~al.,}{Cui et~al.}{2018}]{Cui2018}
Cui W.,  et~al., 2018, \mn@doi [Monthly Notices of the Royal Astronomical
  Society] {10.1093/mnras/sty2111}, 480, 2898

\bibitem[\protect\citeauthoryear{Cybulski, Yun, Fazio  \& Gutermuth}{Cybulski
  et~al.}{2014}]{Cybulski2014}
Cybulski R.,  Yun M.~S.,  Fazio G.~G.,   Gutermuth R.~A.,  2014, \mn@doi
  [Monthly Notices of the Royal Astronomical Society] {10.1093/mnras/stu200},
  439, 3564

\bibitem[\protect\citeauthoryear{Dalton et~al.,}{Dalton
  et~al.}{2012}]{Dalton2012}
Dalton G.,  et~al., 2012, in McLean I.~S.,  Ramsay S.~K.,   Takami H.,  eds,
  Ground-based and Airborne Instrumentation for Astronomy {IV}. {SPIE},
  \mn@doi{10.1117/12.925950}

\bibitem[\protect\citeauthoryear{Darvish, Mobasher, Sobral, Hemmati, Nayyeri
  \& Shivaei}{Darvish et~al.}{2015}]{Darvish2015}
Darvish B.,  Mobasher B.,  Sobral D.,  Hemmati S.,  Nayyeri H.,   Shivaei I.,
  2015, \mn@doi [The Astrophysical Journal] {10.1088/0004-637x/814/2/84}, 814,
  84

\bibitem[\protect\citeauthoryear{Dav{\'{e}}, Angl{\'{e}}s-Alc{\'{a}}zar,
  Narayanan, Li, Rafieferantsoa  \& Appleby}{Dav{\'{e}}
  et~al.}{2019}]{Dave2019}
Dav{\'{e}} R.,  Angl{\'{e}}s-Alc{\'{a}}zar D.,  Narayanan D.,  Li Q.,
  Rafieferantsoa M.~H.,   Appleby S.,  2019, \mn@doi [Monthly Notices of the
  Royal Astronomical Society] {10.1093/mnras/stz937}, 486, 2827

\bibitem[\protect\citeauthoryear{{De Lucia}, {Weinmann}, {Poggianti},
  {Arag{\'o}n-Salamanca}  \& {Zaritsky}}{{De Lucia} et~al.}{2012}]{DeLucia2012}
{De Lucia} G.,  {Weinmann} S.,  {Poggianti} B.~M.,  {Arag{\'o}n-Salamanca} A.,
   {Zaritsky} D.,  2012, \mn@doi [\mnras] {10.1111/j.1365-2966.2012.20983.x},
  \href {http://adsabs.harvard.edu/abs/2012MNRAS.423.1277D} {423, 1277}

\bibitem[\protect\citeauthoryear{Diemer \& Kravtsov}{Diemer \&
  Kravtsov}{2014}]{Diemer2014}
Diemer B.,  Kravtsov A.~V.,  2014, \mn@doi [The Astrophysical Journal]
  {10.1088/0004-637x/789/1/1}, 789, 1

\bibitem[\protect\citeauthoryear{Dolag, Meneghetti, Moscardini, Rasia  \&
  Bonaldi}{Dolag et~al.}{2006}]{Dolag2006}
Dolag K.,  Meneghetti M.,  Moscardini L.,  Rasia E.,   Bonaldi A.,  2006,
  \mn@doi [Monthly Notices of the Royal Astronomical Society]
  {10.1111/j.1365-2966.2006.10511.x}, 370, 656

\bibitem[\protect\citeauthoryear{{Dressler}}{{Dressler}}{1980}]{Dressler1980}
{Dressler} A.,  1980, \mn@doi [\apj] {10.1086/157753}, \href
  {http://adsabs.harvard.edu/abs/1980ApJ...236..351D} {236, 351}

\bibitem[\protect\citeauthoryear{Dressler \& Shectman}{Dressler \&
  Shectman}{1988}]{Dressler1988}
Dressler A.,  Shectman S.~A.,  1988, \mn@doi [The Astronomical Journal]
  {10.1086/114694}, 95, 985

\bibitem[\protect\citeauthoryear{{Dressler}, {Oemler}, {Poggianti}, {Gladders},
  {Abramson}  \& {Vulcani}}{{Dressler} et~al.}{2013}]{dressler13}
{Dressler} A.,  {Oemler} Jr. A.,  {Poggianti} B.~M.,  {Gladders} M.~D.,
  {Abramson} L.,   {Vulcani} B.,  2013, \mn@doi [\apj]
  {10.1088/0004-637X/770/1/62}, \href
  {http://adsabs.harvard.edu/abs/2013ApJ...770...62D} {770, 62}

\bibitem[\protect\citeauthoryear{Durret, Kaastra, Nevalainen, Ohashi  \&
  Werner}{Durret et~al.}{2008}]{Durret2008}
Durret F.,  Kaastra J.~S.,  Nevalainen J.,  Ohashi T.,   Werner N.,  2008,
  \mn@doi [Space Science Reviews] {10.1007/s11214-008-9313-8}, 134, 51

\bibitem[\protect\citeauthoryear{{Eke} et~al.,}{{Eke} et~al.}{2004}]{Eke2004}
{Eke} V.~R.,  et~al., 2004, \mn@doi [\mnras]
  {10.1111/j.1365-2966.2004.07408.x}, \href
  {http://adsabs.harvard.edu/abs/2004MNRAS.348..866E} {348, 866}

\bibitem[\protect\citeauthoryear{Fujita \& Goto}{Fujita \&
  Goto}{2004}]{Fujita2004}
Fujita Y.,  Goto T.,  2004, \mn@doi [Publications of the Astronomical Society
  of Japan] {10.1093/pasj/56.4.621}, 56, 621

\bibitem[\protect\citeauthoryear{{Gao}, {Springel}  \& {White}}{{Gao}
  et~al.}{2005}]{Gao2005}
{Gao} L.,  {Springel} V.,   {White} S.~D.~M.,  2005, \mn@doi [\mnras]
  {10.1111/j.1745-3933.2005.00084.x}, \href
  {http://adsabs.harvard.edu/abs/2005MNRAS.363L..66G} {363, L66}

\bibitem[\protect\citeauthoryear{Geller \& Huchra}{Geller \&
  Huchra}{1983}]{Geller1983}
Geller M.~J.,  Huchra J.~P.,  1983, \mn@doi [The Astrophysical Journal
  Supplement Series] {10.1086/190859}, 52, 61

\bibitem[\protect\citeauthoryear{Gill, Knebe  \& Gibson}{Gill
  et~al.}{2005}]{Gill2005}
Gill S. P.~D.,  Knebe A.,   Gibson B.~K.,  2005, \mn@doi [Monthly Notices of
  the Royal Astronomical Society] {10.1111/j.1365-2966.2004.08562.x}, 356, 1327

\bibitem[\protect\citeauthoryear{Gray et~al.,}{Gray et~al.}{2009}]{Gray2009}
Gray M.~E.,  et~al., 2009, \mn@doi [Monthly Notices of the Royal Astronomical
  Society] {10.1111/j.1365-2966.2008.14259.x}, 393, 1275

\bibitem[\protect\citeauthoryear{Guo, Tempel  \& Libeskind}{Guo
  et~al.}{2015}]{Guo2015}
Guo Q.,  Tempel E.,   Libeskind N.~I.,  2015, \mn@doi [The Astrophysical
  Journal] {10.1088/0004-637x/800/2/112}, 800, 112

\bibitem[\protect\citeauthoryear{Haggar, Gray, Pearce, Knebe, Cui, Mostoghiu
  \& Yepes}{Haggar et~al.}{2020}]{Haggar2020}
Haggar R.,  Gray M.~E.,  Pearce F.~R.,  Knebe A.,  Cui W.,  Mostoghiu R.,
  Yepes G.,  2020, \mn@doi [Monthly Notices of the Royal Astronomical Society]
  {10.1093/mnras/staa273}, 492, 6074

\bibitem[\protect\citeauthoryear{Haines et~al.,}{Haines
  et~al.}{2015}]{Haines2015}
Haines C.~P.,  et~al., 2015, \mn@doi [The Astrophysical Journal]
  {10.1088/0004-637x/806/1/101}, 806, 101

\bibitem[\protect\citeauthoryear{Haines et~al.,}{Haines
  et~al.}{2018a}]{Haines2018a}
Haines C.~P.,  et~al., 2018a, \mn@doi [Monthly Notices of the Royal
  Astronomical Society] {10.1093/mnras/sty651}, 477, 4931

\bibitem[\protect\citeauthoryear{Haines et~al.,}{Haines
  et~al.}{2018b}]{Haines2018}
Haines C.~P.,  et~al., 2018b, \mn@doi [Monthly Notices of the Royal
  Astronomical Society] {10.1093/mnras/sty2338}, 481, 1055

\bibitem[\protect\citeauthoryear{Han, Smith, Choi, Cortese, Catinella, Contini
  \& Yi}{Han et~al.}{2018}]{Han2018}
Han S.,  Smith R.,  Choi H.,  Cortese L.,  Catinella B.,  Contini E.,   Yi
  S.~K.,  2018, \mn@doi [The Astrophysical Journal] {10.3847/1538-4357/aadfe2},
  866, 78

\bibitem[\protect\citeauthoryear{Haynes, Giovanelli  \& Chincarini}{Haynes
  et~al.}{1984}]{Haynes1984}
Haynes M.~P.,  Giovanelli R.,   Chincarini G.~L.,  1984, \mn@doi [Annual Review
  of Astronomy and Astrophysics] {10.1146/annurev.aa.22.090184.002305}, 22, 445

\bibitem[\protect\citeauthoryear{Hess et~al.,}{Hess et~al.}{2018}]{Hess2018}
Hess K.~M.,  et~al., 2018, \mn@doi [Monthly Notices of the Royal Astronomical
  Society] {10.1093/mnras/sty3421}, 484, 2234

\bibitem[\protect\citeauthoryear{Iodice et~al.,}{Iodice
  et~al.}{2019}]{Iodice2019}
Iodice E.,  et~al., 2019, \mn@doi [Astronomy {\&} Astrophysics]
  {10.1051/0004-6361/201935721}, 627, A136

\bibitem[\protect\citeauthoryear{Jaff{\'{e}}, Smith, Candlish, Poggianti, Sheen
   \& Verheijen}{Jaff{\'{e}} et~al.}{2015}]{Jaffe2015}
Jaff{\'{e}} Y.~L.,  Smith R.,  Candlish G.~N.,  Poggianti B.~M.,  Sheen Y.-K.,
   Verheijen M. A.~W.,  2015, \mn@doi [Monthly Notices of the Royal
  Astronomical Society] {10.1093/mnras/stv100}, 448, 1715

\bibitem[\protect\citeauthoryear{Jaff{\'{e}} et~al.,}{Jaff{\'{e}}
  et~al.}{2016}]{Jaffe2016}
Jaff{\'{e}} Y.~L.,  et~al., 2016, \mn@doi [Monthly Notices of the Royal
  Astronomical Society] {10.1093/mnras/stw984}, 461, 1202

\bibitem[\protect\citeauthoryear{Jung, Lee  \& Yi}{Jung
  et~al.}{2014}]{Jung2014}
Jung I.,  Lee J.,   Yi S.~K.,  2014, \mn@doi [The Astrophysical Journal]
  {10.1088/0004-637x/794/1/74}, 794, 74

\bibitem[\protect\citeauthoryear{Jung, Choi, Wong, Kimm, Chung  \& Yi}{Jung
  et~al.}{2018}]{Jung2018}
Jung S.~L.,  Choi H.,  Wong O.~I.,  Kimm T.,  Chung A.,   Yi S.~K.,  2018,
  \mn@doi [The Astrophysical Journal] {10.3847/1538-4357/aadda2}, 865, 156

\bibitem[\protect\citeauthoryear{{Kauffmann}, {White}, {Heckman}, {M{\'e}nard},
  {Brinchmann}, {Charlot}, {Tremonti}  \& {Brinkmann}}{{Kauffmann}
  et~al.}{2004}]{kauffmann04}
{Kauffmann} G.,  {White} S.~D.~M.,  {Heckman} T.~M.,  {M{\'e}nard} B.,
  {Brinchmann} J.,  {Charlot} S.,  {Tremonti} C.,   {Brinkmann} J.,  2004,
  \mn@doi [\mnras] {10.1111/j.1365-2966.2004.08117.x}, \href
  {http://adsabs.harvard.edu/abs/2004MNRAS.353..713K} {353, 713}

\bibitem[\protect\citeauthoryear{Kauffmann, Li  \& Heckman}{Kauffmann
  et~al.}{2010}]{Kauffmann2010}
Kauffmann G.,  Li C.,   Heckman T.~M.,  2010, \mn@doi [Monthly Notices of the
  Royal Astronomical Society] {10.1111/j.1365-2966.2010.17337.x}, 409, 491

\bibitem[\protect\citeauthoryear{Kelkar, Gray, Arag{\'{o}}n-Salamanca, Rudnick,
  Jaff{\'{e}}, Jablonka, Moustakas  \& Milvang-Jensen}{Kelkar
  et~al.}{2019}]{Kelkar2019}
Kelkar K.,  Gray M.~E.,  Arag{\'{o}}n-Salamanca A.,  Rudnick G.,  Jaff{\'{e}}
  Y.~L.,  Jablonka P.,  Moustakas J.,   Milvang-Jensen B.,  2019, \mn@doi
  [Monthly Notices of the Royal Astronomical Society] {10.1093/mnras/stz905},
  486, 868

\bibitem[\protect\citeauthoryear{Kleiner, Pimbblet, Jones, Koribalski  \&
  Serra}{Kleiner et~al.}{2016}]{Kleiner2016}
Kleiner D.,  Pimbblet K.~A.,  Jones D.~H.,  Koribalski B.~S.,   Serra P.,
  2016, \mn@doi [Monthly Notices of the Royal Astronomical Society]
  {10.1093/mnras/stw3328}, p. stw3328

\bibitem[\protect\citeauthoryear{Klypin, Yepes, Gottlöber, Prada  \&
  He{\ss}}{Klypin et~al.}{2016}]{Klypin2016}
Klypin A.,  Yepes G.,  Gottlöber S.,  Prada F.,   He{\ss} S.,  2016, \mn@doi
  [Monthly Notices of the Royal Astronomical Society] {10.1093/mnras/stw248},
  457, 4340

\bibitem[\protect\citeauthoryear{Knebe, Gill, Gibson, Lewis, Ibata  \&
  Dopita}{Knebe et~al.}{2004}]{Knebe2004}
Knebe A.,  Gill S. P.~D.,  Gibson B.~K.,  Lewis G.~F.,  Ibata R.~A.,   Dopita
  M.~A.,  2004, \mn@doi [The Astrophysical Journal] {10.1086/381306}, 603, 7

\bibitem[\protect\citeauthoryear{Knebe et~al.,}{Knebe et~al.}{2020}]{Knebe2020}
Knebe A.,  et~al., 2020, \mn@doi [Monthly Notices of the Royal Astronomical
  Society] {10.1093/mnras/staa1407}, 495, 3002

\bibitem[\protect\citeauthoryear{Kooistra, Silva, Zaroubi, Verheijen, Tempel
  \& Hess}{Kooistra et~al.}{2019}]{Kooistra2019}
Kooistra R.,  Silva M.~B.,  Zaroubi S.,  Verheijen M. A.~W.,  Tempel E.,   Hess
  K.~M.,  2019, \mn@doi [Monthly Notices of the Royal Astronomical Society]
  {10.1093/mnras/stz2677}, 490, 1415

\bibitem[\protect\citeauthoryear{Kraljic et~al.,}{Kraljic
  et~al.}{2018}]{Kraljic2018}
Kraljic K.,  et~al., 2018, \mn@doi [Monthly Notices of the Royal Astronomical
  Society] {10.1093/mnras/sty3216}, 483, 3227

\bibitem[\protect\citeauthoryear{Kuchner et~al.,}{Kuchner
  et~al.}{2020}]{Kuchner2020}
Kuchner U.,  et~al., 2020, \mn@doi [Monthly Notices of the Royal Astronomical
  Society] {10.1093/mnras/staa1083}, 494, 5473

\bibitem[\protect\citeauthoryear{Kuchner et~al.,}{Kuchner
  et~al.}{2021}]{Kuchner2021}
Kuchner U.,  et~al., 2021, \mn@doi [Monthly Notices of the Royal Astronomical
  Society] {10.1093/mnras/stab567}, 503, 2065

\bibitem[\protect\citeauthoryear{Kuutma, Tamm  \& Tempel}{Kuutma
  et~al.}{2017}]{Kuutma2017}
Kuutma T.,  Tamm A.,   Tempel E.,  2017, \mn@doi [Astronomy {\&} Astrophysics]
  {10.1051/0004-6361/201730526}, 600, L6

\bibitem[\protect\citeauthoryear{Laigle et~al.,}{Laigle
  et~al.}{2017}]{Laigle2017}
Laigle C.,  et~al., 2017, \mn@doi [Monthly Notices of the Royal Astronomical
  Society] {10.1093/mnras/stx3055}, 474, 5437

\bibitem[\protect\citeauthoryear{Libeskind et~al.,}{Libeskind
  et~al.}{2017}]{Libeskind2017}
Libeskind N.~I.,  et~al., 2017, \mn@doi [Monthly Notices of the Royal
  Astronomical Society] {10.1093/mnras/stx1976}, 473, 1195

\bibitem[\protect\citeauthoryear{Lisker, Vijayaraghavan, Janz, Gallagher,
  Engler  \& Urich}{Lisker et~al.}{2018}]{Lisker2018}
Lisker T.,  Vijayaraghavan R.,  Janz J.,  Gallagher J.~S.,  Engler C.,   Urich
  L.,  2018, \mn@doi [The Astrophysical Journal] {10.3847/1538-4357/aadae1},
  865, 40

\bibitem[\protect\citeauthoryear{{Mahajan}, {Raychaudhury}  \&
  {Pimbblet}}{{Mahajan} et~al.}{2012}]{Mahajan2012}
{Mahajan} S.,  {Raychaudhury} S.,   {Pimbblet} K.~A.,  2012, \mn@doi [\mnras]
  {10.1111/j.1365-2966.2012.22059.x}, \href
  {http://adsabs.harvard.edu/abs/2012MNRAS.427.1252M} {427, 1252}

\bibitem[\protect\citeauthoryear{Malavasi et~al.,}{Malavasi
  et~al.}{2016}]{Malavasi2016}
Malavasi N.,  et~al., 2016, \mn@doi [Monthly Notices of the Royal Astronomical
  Society] {10.1093/mnras/stw2864}, 465, 3817

\bibitem[\protect\citeauthoryear{Malavasi, Aghanim, Tanimura, Bonjean  \&
  Douspis}{Malavasi et~al.}{2020}]{Malavasi2020}
Malavasi N.,  Aghanim N.,  Tanimura H.,  Bonjean V.,   Douspis M.,  2020,
  \mn@doi [Astronomy {\&} Astrophysics] {10.1051/0004-6361/201936629}, 634, A30

\bibitem[\protect\citeauthoryear{Marinoni, Davis, Newman  \& Coil}{Marinoni
  et~al.}{2002}]{Marinoni2002}
Marinoni C.,  Davis M.,  Newman J.~A.,   Coil A.~L.,  2002, \mn@doi [The
  Astrophysical Journal] {10.1086/343092}, 580, 122

\bibitem[\protect\citeauthoryear{{McGee}, {Balogh}, {Bower}, {Font}  \&
  {McCarthy}}{{McGee} et~al.}{2009}]{McGee2009}
{McGee} S.~L.,  {Balogh} M.~L.,  {Bower} R.~G.,  {Font} A.~S.,   {McCarthy}
  I.~G.,  2009, \mn@doi [\mnras] {10.1111/j.1365-2966.2009.15507.x}, \href
  {http://adsabs.harvard.edu/abs/2009MNRAS.400..937M} {400, 937}

\bibitem[\protect\citeauthoryear{Oemler}{Oemler}{1974}]{Oemler1974}
Oemler A.,  1974, \mn@doi [The Astrophysical Journal] {10.1086/153216}, 194, 1

\bibitem[\protect\citeauthoryear{Paccagnella et~al.,}{Paccagnella
  et~al.}{2017}]{Paccagnella2017}
Paccagnella A.,  et~al., 2017, \mn@doi [The Astrophysical Journal]
  {10.3847/1538-4357/aa64d7}, 838, 148

\bibitem[\protect\citeauthoryear{{Peng}, {Ho}, {Impey}  \& {Rix}}{{Peng}
  et~al.}{2010}]{Peng2010}
{Peng} C.~Y.,  {Ho} L.~C.,  {Impey} C.~D.,   {Rix} H.-W.,  2010, \mn@doi [\aj]
  {10.1088/0004-6256/139/6/2097}, \href
  {http://adsabs.harvard.edu/abs/2010AJ....139.2097P} {139, 2097}

\bibitem[\protect\citeauthoryear{Pimbblet}{Pimbblet}{2010}]{Pimbblet2010}
Pimbblet K.~A.,  2010, \mn@doi [Monthly Notices of the Royal Astronomical
  Society] {10.1111/j.1365-2966.2010.17869.x}, 411, 2637

\bibitem[\protect\citeauthoryear{Poggianti, Smail, Dressler, Couch, Barger,
  Butcher, Ellis  \& Augustus~Oemler}{Poggianti et~al.}{1999}]{Poggianti1999}
Poggianti B.~M.,  Smail I.,  Dressler A.,  Couch W.~J.,  Barger A.~J.,  Butcher
  H.,  Ellis R.~S.,   Augustus~Oemler J.,  1999, \mn@doi [The Astrophysical
  Journal] {10.1086/307322}, 518, 576

\bibitem[\protect\citeauthoryear{Poggianti et~al.,}{Poggianti
  et~al.}{2017}]{Poggianti2017}
Poggianti B.~M.,  et~al., 2017, \mn@doi [The Astrophysical Journal]
  {10.3847/1538-4357/aa78ed}, 844, 48

\bibitem[\protect\citeauthoryear{Porter, Raychaudhury, Pimbblet  \&
  Drinkwater}{Porter et~al.}{2008}]{Porter2008}
Porter S.~C.,  Raychaudhury S.,  Pimbblet K.~A.,   Drinkwater M.~J.,  2008,
  \mn@doi [Monthly Notices of the Royal Astronomical Society]
  {10.1111/j.1365-2966.2008.13388.x}, pp ???--???

\bibitem[\protect\citeauthoryear{{Postman} \& {Geller}}{{Postman} \&
  {Geller}}{1984}]{postman84}
{Postman} M.,  {Geller} M.~J.,  1984, \mn@doi [\apj] {10.1086/162078}, \href
  {http://adsabs.harvard.edu/abs/1984ApJ...281...95P} {281, 95}

\bibitem[\protect\citeauthoryear{Poudel, Heinämäki, Tempel, Einasto, Lietzen
  \& Nurmi}{Poudel et~al.}{2017}]{Poudel2017}
Poudel A.,  Heinämäki P.,  Tempel E.,  Einasto M.,  Lietzen H.,   Nurmi P.,
  2017, \mn@doi [Astronomy {\&} Astrophysics] {10.1051/0004-6361/201629639},
  597, A86

\bibitem[\protect\citeauthoryear{Ramatsoku et~al.,}{Ramatsoku
  et~al.}{2019}]{Ramatsoku2019}
Ramatsoku M.,  et~al., 2019, \mn@doi [Monthly Notices of the Royal Astronomical
  Society] {10.1093/mnras/stz1609}, 487, 4580

\bibitem[\protect\citeauthoryear{Rasia et~al.,}{Rasia et~al.}{2015}]{Rasia2015}
Rasia E.,  et~al., 2015, \mn@doi [The Astrophysical Journal]
  {10.1088/2041-8205/813/1/l17}, 813, L17

\bibitem[\protect\citeauthoryear{Reiprich et~al.,}{Reiprich
  et~al.}{2021}]{Reiprich2021}
Reiprich T.~H.,  et~al., 2021, \mn@doi [Astronomy {\&} Astrophysics]
  {10.1051/0004-6361/202039590}, 647, A2

\bibitem[\protect\citeauthoryear{Rost et~al.,}{Rost et~al.}{2020}]{Rost2020a}
Rost A.,  et~al., 2020, \mn@doi [Monthly Notices of the Royal Astronomical
  Society] {10.1093/mnras/staa3792}, 502, 714

\bibitem[\protect\citeauthoryear{Sarron, Adami, Durret  \& Laigle}{Sarron
  et~al.}{2019}]{Sarron2019}
Sarron F.,  Adami C.,  Durret F.,   Laigle C.,  2019, \mn@doi [Astronomy {\&}
  Astrophysics] {10.1051/0004-6361/201935394}, 632, A49

\bibitem[\protect\citeauthoryear{Sembolini, Yepes, Petris, Gottlöber, Lamagna
  \& Comis}{Sembolini et~al.}{2012}]{Sembolini2012}
Sembolini F.,  Yepes G.,  Petris M.~D.,  Gottlöber S.,  Lamagna L.,   Comis
  B.,  2012, \mn@doi [Monthly Notices of the Royal Astronomical Society]
  {10.1093/mnras/sts339}, 429, 323

\bibitem[\protect\citeauthoryear{Sousbie}{Sousbie}{2011}]{Sousbie2011}
Sousbie T.,  2011, \mn@doi [Monthly Notices of the Royal Astronomical Society]
  {10.1111/j.1365-2966.2011.18394.x}, 414, 350

\bibitem[\protect\citeauthoryear{Tanimura, Aghanim, Douspis, Beelen  \&
  Bonjean}{Tanimura et~al.}{2019}]{Tanimura2019a}
Tanimura H.,  Aghanim N.,  Douspis M.,  Beelen A.,   Bonjean V.,  2019, \mn@doi
  [Astronomy {\&} Astrophysics] {10.1051/0004-6361/201833413}, 625, A67

\bibitem[\protect\citeauthoryear{Tempel, Stoica, Mart{\'{\i}}nez, Liivamägi,
  Castellan  \& Saar}{Tempel et~al.}{2014}]{Tempel2014}
Tempel E.,  Stoica R.~S.,  Mart{\'{\i}}nez V.~J.,  Liivamägi L.~J.,  Castellan
  G.,   Saar E.,  2014, \mn@doi [Monthly Notices of the Royal Astronomical
  Society] {10.1093/mnras/stt2454}, 438, 3465

\bibitem[\protect\citeauthoryear{Umehata et~al.,}{Umehata
  et~al.}{2019}]{Umehata2019}
Umehata H.,  et~al., 2019, \mn@doi [Science] {10.1126/science.aaw5949}, 366, 97

\bibitem[\protect\citeauthoryear{Vijayaraghavan \& Ricker}{Vijayaraghavan \&
  Ricker}{2013}]{Vijayaraghavan2013}
Vijayaraghavan R.,  Ricker P.~M.,  2013, \mn@doi [Monthly Notices of the Royal
  Astronomical Society] {10.1093/mnras/stt1485}, 435, 2713

\bibitem[\protect\citeauthoryear{Vulcani et~al.,}{Vulcani
  et~al.}{2019}]{Vulcani2019}
Vulcani B.,  et~al., 2019, \mn@doi [Monthly Notices of the Royal Astronomical
  Society] {10.1093/mnras/stz1399}, 487, 2278

\bibitem[\protect\citeauthoryear{{Wetzel}, {Tinker}, {Conroy}  \& {van den
  Bosch}}{{Wetzel} et~al.}{2013}]{wetzel13}
{Wetzel} A.~R.,  {Tinker} J.~L.,  {Conroy} C.,   {van den Bosch} F.~C.,  2013,
  \mn@doi [\mnras] {10.1093/mnras/stt469}, \href
  {http://adsabs.harvard.edu/abs/2013MNRAS.432..336W} {432, 336}

\bibitem[\protect\citeauthoryear{White, Cohn  \& Smit}{White
  et~al.}{2010}]{White2010}
White M.,  Cohn J.~D.,   Smit R.,  2010, \mn@doi [Monthly Notices of the Royal
  Astronomical Society] {10.1111/j.1365-2966.2010.17248.x}, 408, 1818

\bibitem[\protect\citeauthoryear{Winkel, Pasquali, Kraljic, Smith, Gallazzi  \&
  Jackson}{Winkel et~al.}{2021}]{Winkel2021}
Winkel N.,  Pasquali A.,  Kraljic K.,  Smith R.,  Gallazzi A.,   Jackson T.~M.,
   2021, \mn@doi [Monthly Notices of the Royal Astronomical Society]
  {10.1093/mnras/stab1562}, 505, 4920

\bibitem[\protect\citeauthoryear{Yang, Mo, Jing  \& van~den Bosch}{Yang
  et~al.}{2005}]{Yang2005}
Yang X.,  Mo H.~J.,  Jing Y.~P.,   van~den Bosch F.~C.,  2005, \mn@doi [Monthly
  Notices of the Royal Astronomical Society]
  {10.1111/j.1365-2966.2005.08801.x}, 358, 217

\bibitem[\protect\citeauthoryear{Zabludoff \& Mulchaey}{Zabludoff \&
  Mulchaey}{1998}]{Zabludoff1998}
Zabludoff A.~I.,  Mulchaey J.~S.,  1998, \mn@doi [The Astrophysical Journal]
  {10.1086/305355}, 496, 39

\bibitem[\protect\citeauthoryear{Zel'dovich}{Zel'dovich}{1970}]{Zeldovich1970}
Zel'dovich Y.~B.,  1970, \aap, 5

\bibitem[\protect\citeauthoryear{van Haarlem \& van~de Weygaert}{van Haarlem \&
  van~de Weygaert}{1993}]{Haarlem1993}
van Haarlem M.,  van~de Weygaert R.,  1993, \mn@doi [The Astrophysical Journal]
  {10.1086/173416}, 418, 544

\bibitem[\protect\citeauthoryear{van~de Weygaert, Shandarin, Saar  \&
  Einasto}{van~de Weygaert et~al.}{2014}]{vandeWeygeartProceedings2014}
van~de Weygaert R.,  Shandarin S.,  Saar E.,   Einasto J.,  eds, 2014, {IAU}
  S308 The Zeldovich Universe: Genesis and Growth of the Cosmic Web  IAU
  symposium proceedings series Vol. 11.
Cambridge University Press ({CUP})

\makeatother
\end{thebibliography}








\bsp	
\label{lastpage}
\end{document}